\documentclass[final]{elsarticle}

\usepackage{lineno,hyperref}
\modulolinenumbers[5]

\journal{Journal of Linear Algebra and its Applications}

\renewcommand\Re{\operatorname{Re}}

\usepackage{amsmath, amssymb, accents}
\usepackage{mathrsfs}
\usepackage{hyperref}
\usepackage{color}
\usepackage{subfig}
\usepackage{multirow}

\newtheorem{theorem}{Theorem}
\newtheorem{definition}[theorem]{Definition} 
\newtheorem{proposition}[theorem]{Proposition}
\newtheorem{remark}[theorem]{Remark}

\def\Diag{\operatorname{Diag}}
\def\diag{\operatorname{diag}}
\def\vec{\operatorname{vec}}
\def\argmin#1{\operatornamewithlimits{arg\, min}_{#1}}
\def\tr{\operatorname{tr}}

\def\holmat#1{\accentset{\circ}{#1}}

\if@mathematic
   \def\bd#1{\ensuremath{\mathchoice
                     {\mbox{\boldmath$\displaystyle\mathbf{#1}$}}
                     {\mbox{\boldmath$\textstyle\mathbf{#1}$}}
                     {\mbox{\boldmath$\scriptstyle\mathbf{#1}$}}
                     {\mbox{\boldmath$\scriptscriptstyle\mathbf{#1}$}}}}
\else
   \def\bd#1{\ensuremath{\mathchoice
                     {\mbox{\boldmath$\displaystyle#1$}}
                     {\mbox{\boldmath$\textstyle#1$}}
                     {\mbox{\boldmath$\scriptstyle#1$}}
                     {\mbox{\boldmath$\scriptscriptstyle#1$}}}}
\fi

\begin{document}

\begin{frontmatter}

\title{Diagonality Measures of Hermitian Positive-Definite Matrices with Application to the Approximate Joint Diagonalization Problem}


\author[enitaddress]{Khaled Alyani}
\ead{khaled.alyani@yahoo.de}

\author[gipsaaddress]{Marco Congedo}
\ead{marco.congedo@gipsa-lab.grenoble-inp.fr}

\author[enitaddress]{Maher Moakher\corref{mycorrespondingauthor}}
\ead[url]{http://sites.google.com/site/moakhersite/}
\ead{maher.moakher@gmail.com}
\cortext[mycorrespondingauthor]{Corresponding author}

\address[enitaddress]{LAMSIN-ENIT, University of Tunis El Manar, B.P. 37, 1002 Tunis-Belv\'ed\`ere, Tunisia}

\address[gipsaaddress]{GIPSA-lab, CNRS and Grenoble University, Grenoble, France}

\begin{abstract}
In this paper, we introduce properly-invariant diagonality measures of Hermitian positive-definite matrices. These diagonality measures are defined as distances or divergences between a given positive-definite matrix and its diagonal part. We then give closed-form expressions of these diagonality measures and discuss their invariance properties. The diagonality measure based on the log-determinant $\alpha$-divergence is general enough as it includes a diagonality criterion used by the signal processing community as a special case. These diagonality measures are then used to formulate minimization problems for finding the approximate joint diagonalizer of a given set of Hermitian positive-definite matrices. Numerical computations based on a modified Newton method are presented and commented.
\end{abstract}


\end{frontmatter}


\section{Introduction}

In the last three decades, the problem of approximate joint diagonalization (AJD) of a collection of Hermitian positive-definite matrices has attracted an increasing attention from researchers within the statistics, signal processing and applied mathematics communities. In statistics, its is used to solve the common principal component (CPC) problem \cite{Flury86,Flury84}. In signal processing it is a fundamental tool for blind source separation (BSS) \cite{Comon10} and blind beamforming \cite{cardoso1993, veen1998}. Applied mathematicians and numerical analysts are concerned with efficient numerical algorithms for solving this problem \cite{Joho08,pham01,cardoso1996jacobi,mesloub,tichavsky09,yeredor2002non,yeredor2004approximate,ziehe2004fast}.

Given a set of $n \times n$ Hermitian positive-definite matrices $\bd{M}_k$, $k=1,\ldots, K$ that are supposed to be diagonalized under congruence by a common invertible matrix $\bd{C}$, i.e., such that
\[
\bd{D}_k = \bd{C} \bd{M}_k \bd{C}^H, \quad k=1, \ldots, K,
\]  
where $\bd{D}_k$ are diagonal matrices, or equivalently, such that $\bd{C}^{-1} \bd{D}_k \bd{C}^{-H}=\bd{M}_k$, $k=1, \ldots, K$, the problem of exact joint diagonalizer is to find a common diagonalizer for the given $\bd{M}_k$, $k=1,\ldots, K$. We note that the joint diagonalizer can only be determined up to a row-wise permutation and scaling. In fact, if $\bd{C}$ is a common diagonalizer of $\bd{M}_k$, $k=1,\ldots, K$ then so is $\bd{P}\bd{D}\bd{C}$ for any invertible diagonal matrix $\bd{D}$ and any permutation matrix $\bd{P}$.

In practice, due to the presence of noise and measurement errors, the given (observed) matrices $\bd{M}_k$, $k=1,\ldots, K$ can not be simultaneously diagonalized for $K > 2$. The problem of approximate joint diagonalizer for the given $\bd{M}_k$, $k=1,\ldots, K$ is then to find an invertible matrix $\bd{C}$ such that $\bd{C} \bd{M}_k \bd{C}^{H}$, $k=1, \ldots, K$ are as close to diagonal matrices as possible \cite{pham01,cardoso1996jacobi,Joho08,mesloub,tichavsky09,yeredor2002non,yeredor2004approximate,ziehe2004fast}. Hence, one has to define a measure of diagonality and then to devise algorithms that minimize a cost function which is the sum of the diagonality measures of all $\bd{C} \bd{M}_{k} \bd{C}^{H}$.

Since the joint diagonalizer is not unique, but arbitrary with respect to permutations and positive diagonal scalings, it is desirable that the cost function used for computing the approximate joint diagonalizer be invariant with respect to permutations and positive diagonal scalings. The overwhelming majority of the proposed approximate joint diagonalization algorithms use as cost function the Frobenius off squared norm, which is not invariant by diagonal scaling. Furthermore, this criterion is not specific to positive Hermitian matrices. The reason why this cost function has been heavily employed is because it allows simple optimization schemes. On the other hand, Pham \cite{pham01} used a cost function which is invariant with respect to diagonal scalings and permutations and is specifically conceived for Hermitian positive-definite matrices. He used a diagonality measure derived from the Kullback-Leibler divergence to minimize an unconstrained optimization problem, in contrast to most joint diagonalization algorithms, which minimize a constrained optimization problem. We will propose a new cost function, based on the log-det $\alpha$-divergence, that measures the degree of joint diagonalization of a set of Hermitian positive-definite matrices. This cost function includes some of the cost functions used in the literature as special cases. To the best of our knowledge, this paper is the first attempt to tackle the approximate joint diagonalizer problem from the Riemannian and information geometries point of view.
We then describe convergent numerical algorithms for minimizing this cost function and hence finding a single matrix $\bd{C}$ that approximately joint diagonalizes the given set of matrices.

The remainder of this paper is organized as follows. In Section~\ref{sec:back}, we set notations and gather necessary background materials from matrix analysis, differential geometry, optimization on manifolds and information theory that will be used in the sequel. In Section~\ref{sec:diag}, and as a first step, we start by introducing properly-invariant diagonality measures as distances or divergences between a given positive-definite matrix and its diagonal part. Properly invariant means that all desirable invariance properties (which will be made precise) that one would expect of such measures are satisfied. For generality, we have chosen to work with Hermitian positive-definite matrices. The results in this paper remain valid for symmetric positive-definite matrices, all what is needed is to consider that all matrices are real. We then give closed-form expressions of these diagonality measures and discuss their properties. Then, in Section~\ref{sec:tdiag}, we define the ``true'' diagonality measures as the distances (or divergences) between a given positive-definite matrix and its closest, with respect to the distance (divergence), diagonal matrix. Unfortunately, not all of them can be given in closed-form. For those which can not be given explicitly, we give the nonlinear matrix equation that characterize them. In Section~\ref{sec:jad} application to the problem of approximate joint diagonalization of a finite collection of Hermitian positive-definite matrices is discussed and some computations of the approximate joint diagonalizer based on these diagonality measures are presented.

\section{Notations and background materials} \label{sec:back}

Let $\mathcal{M}_n(\mathbb{C})$ be the set of $n \times n$ complex matrices and ${\rm GL}_n(\mathbb{C})$ be its subset containing only nonsingular matrices. ${\rm GL}_n(\mathbb{C})$ is a Lie group, i.e., a group which is also a differentiable manifold and for which the operations of group multiplication and inverse are smooth. The tangent space at the identity is called the corresponding Lie algebra and denoted by $\mathfrak{gl}_n(\mathbb{C})$. It is the space of all linear transformations in $\mathbb{C}^n$, i.e., $\mathcal{M}_n(\mathbb{C})$. In $\mathcal{M}_n(\mathbb{C})$ we shall use the Euclidean inner product, known as the Frobenius inner product, defined by 
\[
\langle \bd{A}, \bd{B} \rangle_F := \tr(\bd{B}^H \bd{A}),
\] 
where $\tr(\cdot)$ stands for the trace and the superscript $^H$ denotes the conjugate transpose. The associated norm $\| \bd{A} \|_F = \langle \bd{A}, \bd{A} \rangle_F^{1/2}$ is used to define the Euclidean distance on $\mathcal{M}_n(\mathbb{C})$,
\[
d_F (\bd{A},\bd{B}) := \| \bd{A} - \bd{B} \|_F.
\]

\subsection{The $\Diag$ operator}

We will denote by $\Diag$ the linear operator, from $\mathcal{M}_n(\mathbb{C})$
into itself, which assigns to each matrix $\bd{A}$ the diagonal matrix $\Diag (\bd{A})$ whose diagonal entries are $a_{kk}$, the diagonal entries of  $\bd{A}$. We note that $\Diag(\bd{A})$ can be expressed as $\Diag(\bd{A}) = \bd{A} \circ \bd{I} = \bd{I} \circ \bd{A}$, where $\circ$ represents the Hadamard product (entry-wise product). It is also known as the Schur product.  
Then, if we represent linear maps from $\mathcal{M}_n(\mathbb{C})$ into itself by its $n^2 \times n^2$ matrix in the Canonical basis, it is easy to see that the differential of $\Diag(\bd{A})$ with respect to $\bd{A}$ is given by
\[
D \Diag (\bd{A}) = \diag (\vec (\bd{I})),
\]
where $\vec (\cdot)$ stands for the operator that transforms an $n \times n$ complex-valued matrix into a vector in $\mathbb{C}^{n^2}$ by stacking the columns of the matrix one underneath the other, and $\diag (\cdot)$ is the operator that assigns to each vector in $\mathbb{C}^{m}$ an $m \times m$ diagonal matrix with the elements of the vector on its main diagonal. 

Let us define the diagonal unitary matrix $\bd{U}_\theta = \diag(e^{i\theta}, e^{2i\theta}, \ldots, e^{n i \theta})$, where $\theta \in [0, 2\pi)$. Then, it is interesting to observe that $\Diag(\bd{A})$ can be represented as the continuous average of rotated versions of $\bd{A}$ \cite{Bhatia00,Bhatia89}
\[
\Diag(\bd{A}) = \frac1{2\pi} \int_{0}^{2\pi} \bd{U}_\theta \bd{A} \bd{U}_\theta^{H} \; d\theta,   
\]
or, as the discrete average of rotated versions of $\bd{A}$
\[
\Diag(\bd{A}) = \frac1n \sum_{k=1}^{n} \bd{U}_{\theta_k} \bd{A} \bd{U}_{\theta_k}^{H},   \quad \text{where }\theta_k = \frac{2(k-1)\pi}n.
\]
So that the differential of $\Diag(\bd{A})$ with respect to $\bd{A}$ can also be written as
\[
D \Diag (\bd{A}) = \frac1{2\pi} \int_{0}^{2\pi} \bd{U}_\theta \otimes \bd{U}_\theta \; d\theta = \frac1n \sum_{k=1}^{n} \bd{U}_{\theta_k} \otimes \bd{U}_{\theta_k},
\]
where $\otimes$ stands for the Kronecker product.

\subsection{Exponential and logarithms} 

The exponential of a matrix in $\mathfrak{gl}_n(\mathbb{C})$ is given, as usual, by the convergent series
\[
\exp \bd{A} = \sum_{k=0}^\infty \frac1{k!} \bd{A}^k.
\]
We remark that the product of the exponentials of two matrices $\bd{A}$ and $\bd{B}$ is equal to
$\exp( \bd{A} + \bd{B})$ only when $\bd{A}$ and $\bd{B}$ commute.

Logarithms of $\bd{A}$ in $\mathcal{M}_n(\mathbb{C})$ are solutions of the matrix equation $\exp \bd{X} = \bd{A}$.
When $\bd{A}$ does not have eigenvalues on the closed negative real line, there exists a unique logarithm, called the principal logarithm and denoted by $\log \bd{A}$, whose spectrum lies in the infinite strip 
$\left\{z \in \mathbb{C}\ : \ -\pi < \mathrm{Im}(z) < \pi \right\}$,
of the complex plane. Furthermore, if for a given matrix norm $\| \cdot \|$ we have $\| \bd{A} \| < 1$, then the series $\sum_{k=1}^\infty \frac{(-1)^{k-1}}{k} \bd{A}^k$
converges to $\log (\bd{I} + \bd{A})$, where $\bd{I}$ denotes the identity transformation in $\mathbb{C}^n$. Therefore, one can write
\[
\log (\bd{I} + \bd{A}) = \sum_{k=1}^\infty \frac{(-1)^{k-1}}{k} \bd{A}^k.
\]
We note that, in general, $\log(\bd{A} \bd{B}) \neq \log \bd{A} + \log \bd{B}$. Finally, we recall the important fact:
\[
f(\bd{C}^{-1} \bd{B} \bd{C}) = \bd{C}^{-1} f(\bd{B}) \bd{C},
\]
for any invertible matrix $\bd{C}$ and any analytic function $f$, such as $\exp$ and $\log$.

\subsection{Geodesic convexity, gradient and Hessian} 

For a Riemannian manifold $\mathscr{M}$ we denote by $T_x\mathscr{M}$ the tangent space at $x \in \mathscr{M}$ and by $\exp_x : T_x\mathscr{M} \to \mathscr{M}$ the exponential mapping at $x$, i.e. $\exp_x(v) := \gamma(1)$ where $\gamma$ is the unique geodesic with $\gamma(0) = x$ and $\gamma'(0) = v$. 

A subset $\mathscr{A}$ of a Riemannian manifold $\mathscr{M}$ is said to be convex if the shortest geodesic curve between any two points $x$ and $y$ in $\mathscr{A}$ is unique in $\mathscr{M}$ and lies entirely in $\mathscr{A}$. A real-valued function defined on a convex subset $\mathscr{A}$ of $\mathscr{M}$ is said to be convex if its restriction to any geodesic path is convex, i.e., if $t \mapsto \hat{f}(t) := f(\exp_x(tu))$ is convex over its domain for all $x \in \mathscr{M}$ and $u \in T_x(\mathscr{M})$.

Let $f$ be a real-valued function defined on a Riemannian manifold $\mathscr{M}$. If $f$ is differentiable then the gradient $\nabla f$ is the unique tangent vector at $x$ such that for any direction $u$ in the tangent space to $\mathscr{M}$ at $x$, we have
\[
\langle u, \nabla f \rangle = \left. \frac{d}{dt}\right|_{t=0} f(\exp_x(tu)),
\]
where $\langle \cdot, \cdot \rangle$ denotes the Riemannian inner product on the tangent space. If $f$ is twice differentiable, then the Hessian $\nabla^2 f$ of $f$ is given by
\[
\langle u, \nabla^2 f v \rangle = \left. \frac{\partial^2}{\partial t \partial s}\right|_{t=0, s=0} f(\exp_x(tu + s v)).
\]

The second-order Taylor expansion of the function $f$ at $x \in \mathscr{M}$ is therefore given by
\[
f(\exp_x(v)) = f(x) + \langle \nabla f, v \rangle +\tfrac12 \langle v, \nabla^2 f v \rangle + O(|v|^3). 
\]

\subsection{The cone of Hermitian positive-definite matrices} 

We denote by
\[
\mathscr{H}_n := \left\{ \bd{A} \in \mathcal{M}_n(\mathbb{C}), \ \bd{A}^H = \bd{A} \right\},
\]
the $n^2$-dimensional vector space of all $n \times n$ Hermitian matrices and denote by
\[
\mathscr{H}^{++}_n := \left\{ \bd{A} \in \mathscr{H}_n, \ \bd{A} > 0 \right\},
\]
the set of all $n \times n$ Hermitian positive-definite matrices. Here $\bd{A} > 0$ means that the quadratic form $\bd{x}^H \bd{A} \bd{x}$ is positive for all $\bd{x} \in \mathbb{C}^n\setminus \{ \bd{0} \}$. It is well known that $\mathscr{H}^{++}_n$ is an open convex cone; i.e., if $\bd{P}$ and $\bd{Q}$ are in $\mathscr{H}^{++}_n$, so is $\bd{P} + t \bd{Q}$ for any $t > 0$. Its closure is the set $\mathscr{H}^{+}_n$ of all $n \times n$ Hermitian positive semi-definite matrices and its boundary $\partial \mathscr{H}^{++}_n$ is the set of all $n \times n$ Hermitian positive semi-definite singular matrices.

We recall that the exponential map from $\mathscr{H}_n$ to $\mathscr{H}^{++}_n$ is one-to-one and onto. In other words, the exponential of any Hermitian matrix is a Hermitian positive-definite matrix, and the inverse of the exponential (i.e., the principal logarithm) of any Hermitian positive-definite matrix is a Hermitian matrix.

As $\mathscr{H}^{++}_n$ is an open subset of $\mathscr{H}_n$, for each $\bd{P} \in\mathscr{H}^{++}_n$ we identify the set $T_{\bd{P}}$ of
tangent vectors to $\mathscr{H}^{++}_n$ at $\bd{P}$ with $\mathscr{H}_n$. On the tangent space at $\bd{P}$ we define the positive-definite inner product and corresponding norm \cite{Moakher05,bhatia07}
\begin{equation} \label{eq:ip}
\langle \bd{A}, \bd{B} \rangle_{\bd{P}} = \tr( \bd{P}^{-1} \bd{A} \bd{P}^{-1} \bd{B}), \qquad \| \bd{A} \|_{\bd{P}} = \langle \bd{A}, \bd{A} \rangle^{1/2}_{\bd{P}},
\end{equation}
that smoothly depend on the base point $\bd{P}$. The positive definiteness is a consequence of the positive definiteness of the Frobenius inner product for
\[
\langle \bd{A}, \bd{A} \rangle_{\bd{P}} = \tr( \bd{P}^{-1/2} \bd{A} \bd{P}^{-1/2} \bd{P}^{-1/2} \bd{A} \bd{P}^{-1/2}) = \| \bd{P}^{-1/2} \bd{A} \bd{P}^{-1/2} \|_F^2.
\]

Let $[a, b]$ be a closed interval in $\mathbb{R}$, and let $\bd{\Gamma} : [a, b] \to \mathscr{H}^{++}_n$ be a sufficiently smooth curve in $\mathscr{H}^{++}_n$. We define the length of $\bd{\Gamma}$ by
\[
\mathcal{L}(\bd{\Gamma}) := \int_a^b  \| \dot{\bd{\Gamma}}(t) \|_{\bd{\Gamma(t)}} dt.
\]

We note that the length $\mathcal{L}(\bd{\Gamma})$ is invariant under congruent transformations, i.e., $\bd{\Gamma} \mapsto \bd{C} \bd{\Gamma} \bd{C}^H$, where $\bd{C}$ is any fixed element of ${\rm GL}_n(\mathbb{C})$. As $\frac{d \bd{\Gamma}^{-1}}{dt} = - \bd{\Gamma}^{-1} \frac{d \bd{\Gamma}}{dt} \bd{\Gamma}^{-1}$, one can
readily see that this length is also invariant under inversion.

The distance between two matrices $\bd{A}$ and $\bd{B}$ in $\mathscr{H}^{++}_n$, considered as a differentiable manifold, is the infimum of lengths of curves connecting them:
\[
d_{\mathscr{H}^{++}_n}(\bd{A}, \bd{B}) := \inf \left\{ \mathcal{L}(\bd{\Gamma}) | \bd{\Gamma} : [a, b] \to \mathscr{H}^{++}_n \text{ with } \bd{\Gamma}(a) = \bd{A}, \bd{\Gamma}(b) = \bd{B} \right\}.
\]
This metric makes $\mathscr{H}^{++}_n$ a Riemannian manifold of dimension $n^2$.

The geodesic emanating from $\bd{I}$ in the direction of $\bd{S}$, a Hermitian matrix in the tangent space, is given explicitly by $\exp(t \bd{S})$. Using invariance under congruent transformations, the geodesic $\bd{P}(t)$ such that $\bd{P}(0) = \bd{P}$ and $\dot{\bd{P}}(0) = \bd{S}$ is therefore given by
\[
\bd{P}(t) = \bd{P}^{1/2} \exp(t \bd{P}^{-1/2} \bd{S} \bd{P}^{-1/2}) \bd{P}^{1/2}.
\]
It follows that the Riemannian distance between $\bd{P}_1$ and $\bd{P}_2$ in $\mathscr{H}^{++}_n$ is
\begin{equation} \label{eq:RiemDist}
d_{\mathscr{H}^{++}_n}(\bd{P}_1, \bd{P}_2) = \| \log( \bd{P}_1^{-1/2} \bd{P}_2 \bd{P}_1^{-1/2}) \|_F = \left[ \sum_{i=1}^n \log^2 \lambda_i \right]^{1/2},
\end{equation}
where $\lambda_i$, $i = 1, \ldots, n$, are the (positive) eigenvalues of $\bd{P}_1^{-1} \bd{P}_2$. Even though in general $\bd{P}_1^{-1} \bd{P}_2$ is not Hermitian, its eigenvalues are real and positive. This can be seen by noting that $\bd{P}_1^{-1} \bd{P}_2$ is similar to the Hermitian positive-definite matrix $\bd{P}_1^{-1/2} \bd{P}_2 \bd{P}_1^{-1/2}$. It is important to note here that the real-valued function defined on $\mathscr{H}^{++}_n$ by $\bd{P} \mapsto d_{\mathscr{H}^{++}_n}(\bd{P}, \bd{S})$, where $\bd{S} \in \mathscr{H}^{++}_n$ is fixed, is (geodesically) convex. We note in passing that $\mathscr{H}^{++}_n$ is a homogeneous space of the Lie group ${\rm GL}_n(\mathbb{C})$ (by identifying $\mathscr{H}^{++}_n$ with the quotient ${\rm GL}_n(\mathbb{C})/U_n(\mathbb{C})$). It is also a symmetric space of noncompact type \cite{terras}.

It is important here to note that the subset $\mathscr{D}_n^{++}$ of diagonal positive-definite matrices is a totally geodesic submanifold of $\mathscr{H}_n^{++}$, which is geodesically convex and closed with respect to the Riemannian metric defined by the inner product (\ref{eq:ip})$_1$.

\subsection{Divergence functions}

We give a formal definition of a divergence function on a Riemannian manifold.
\begin{definition}
A divergence function over a Riemannian manifold $\mathscr{M}$ is a real-valued function $J$ on the Cartesian product manifold $\mathscr{M} \times \mathscr{M}$ which satisfies the following conditions \cite{Amari12}:
\begin{itemize}
\item[i)] Positive definiteness: $J(x, y) \ge 0$ for all $x$ and $y$ in $\mathscr{M}$ with equality if and only if $x = y$. 

\item[ii)] Differentiability: $J(x, y)$ is twice differentiable with respect to $x$ and its Hessian with respect to $x$, when evaluated at $y=x$, is positive definite.
\end{itemize}
\end{definition}

We note that the divergence is almost a distance function except that it needs not to be symmetric with respect to its arguments nor to satisfy the triangle inequality. For instance, the square of a distance function is a (symmetric) divergence function. In some respects, a divergence function is a generalization of squared distances and has the physical dimension (i.e., unit) of squared distance.

Let $f: \Omega \to \mathbb{R}$ be a twice differentiable and strictly convex function defined on a closed convex set $\Omega$ of a vector space $V$. We give below two examples of systematic ways of constructing divergence functions from $f$.

\begin{enumerate}
\item The function defined by
\begin{equation} \label{eq:breg-div}
B_f (x, y) = f (x) - f (y) - \langle \nabla f (y), x - y \rangle,
\end{equation}
is a divergence function on $\Omega$ called the Bregman divergence \cite{Bregman67}. Here $\langle \cdot, \cdot \rangle$ denotes the inner product in $V$.

\item For a real parameter $\alpha$ such that $| \alpha | <1$, the functions 
\begin{equation} \label{eq:alpha-div}
D_f^{\alpha}(x, y) = \tfrac{4}{1 - \alpha^2} \left[ \tfrac{1 - \alpha}{2} f(x) + \tfrac{1 + \alpha}{2} f(y) - f(\tfrac{1 - \alpha}{2} x + \tfrac{1 + \alpha}{2} y) \right],
\end{equation}
defines a one-parameter family of divergence functions called $\alpha$-divergence functions \cite{Zhang04}. Furthermore, by taking limits in (\ref{eq:alpha-div}) we have
\begin{align*}
& D_f^{1} (x, y) := \lim_{\alpha \to 1^-} D_f^{\alpha}(x, y) = B_f (x, y), \\
& D_f^{-1} (x, y) := \lim_{\alpha \to -1^+} D_f^{\alpha}(x, y) = B_f (y, x).
\end{align*}
\end{enumerate}

The positive definiteness of the functions (\ref{eq:breg-div}) and (\ref{eq:alpha-div}) is guaranteed by the strict convexity of $f$. Furthermore, we note that the Hessian with respect to $x$ when evaluated at $y=x$ for both functions (\ref{eq:breg-div}) and (\ref{eq:alpha-div}) is equal to the Hessian of $f$ at $x$ which is positive definite by the strict convexity hypothesis on $f$.

Now, if we consider the logarithmic-barrier function $f(\bd{X}) = - \log \det \bd{X}$, which is a strictly convex function defined on the convex subset $\mathscr{H}_n^{++}$ of the vector space $\mathscr{H}_n$, we obtain for $-1 \le \alpha \le 1$, the one-parameter family of divergence functions on the space of Hermitian positive-definite matrices called log-det $\alpha$-divergence functions \cite{Chebbi12}
\begin{equation} \label{eq:mat-alpha-div}
\begin{cases}
D_{LD}^{\alpha} (\bd{A}, \bd{B}) = \tfrac{4}{1 - \alpha^2} \log \frac{\det \left( \tfrac{1 - \alpha}{2} \bd{A} + \tfrac{1 + \alpha}{2} \bd{B} \right)}{\big(\det \bd{A}\big)^{\tfrac{1 - \alpha}{2}} \big(\det \bd{B} \big)^{\tfrac{1 + \alpha}{2}} }, & -1 < \alpha < 1, \\[.5cm]
D_{LD}^{-1} (\bd{A}, \bd{B}) = \tr( \bd{A}^{-1} \bd{B} - \bd{I}) - \log \det( \bd{A}^{-1} \bd{B}), \\[.5cm]
D_{LD}^{1} (\bd{A}, \bd{B}) = \tr( \bd{B}^{-1} \bd{A} - \bd{I}) - \log \det( \bd{B}^{-1} \bd{A}).
\end{cases}
\end{equation}

\subsection{Projection onto a convex set}

Let $\mathscr{S}$ be a nonempty closed convex subset of a Riemannian manifold $\mathscr{M}$. If $d(\cdot, \cdot)$ is a distance function on $\mathscr{M}$, then for any $x \in \mathscr{M}$ there exists a unique element $x_d \in \mathscr{S}$ such that $d(x,x_d) \le d(x,y)$ for all $y \in \mathscr{S}$. The element $x_d$ is called the metric projection of $x$ onto $\mathscr{S}$.

Similarly, if $D(\cdot, \cdot)$ is a divergence function on $\mathscr{M}$, then for any $x \in \mathscr{M}$ there exists a unique element $x_D \in \mathscr{S}$ such that $D(x,x_D) \le D(x,y)$ for all $y \in \mathscr{S}$ \cite{amari09}. The element $x_D$ is called the divergence projection of $x$ onto $\mathscr{S}$.

\section{Properly invariant diagonality measures} \label{sec:diag}

We define a measure of diagonality of a matrix as a non-negative quantity 
that vanishes only for diagonal matrices. One way to construct such measures 
is to use a distance (or a divergence) function and define the measure of
diagonality as the squared distance (or the divergence) between the given 
matrix and its diagonal part.
The minimal set of invariance properties that we would expect from a 
diagonality measure $\mathscr{D}(\cdot)$ are the following:
\begin{enumerate}
\item Invariance under congruence transformations by permutation matrices:
$$
\mathscr{D}(\bd{P} \bd{A} \bd{P}) = \mathscr{D}(\bd{A}), \text{ for all 
permutation matrices } \bd{P}.
$$

\item Invariance under congruence transformations by invertible diagonal 
matrices:
$$
\mathscr{D}(\bd{D} \bd{A} \bd{D}) = \mathscr{D}(\bd{A}), \text{ for all 
invertible diagonal matrices } \bd{D}.
$$
\end{enumerate}
Note that Property~2 implies invariance under positive scaling, 
i.e., $\mathscr{D}(\mu \bd{A}) = \mathscr{D}(\bd{A})$, $\forall \mu>0$.

The simplest diagonality measure of a given matrix $\bd{A}$ could be obtained 
by half the squared norm of its off-diagonal part $\bd{A} - \Diag \bd{A}$.
For example, if we consider the Euclidean distance induced from the matrix 
Frobenius inner product, then the Frobenius measure of diagonality of a 
Hermitian positive-definite matrix $\bd{A}$ is defined as half the square 
of the Frobenius distance between $\bd{A}$ and $\Diag \bd{A}$, i.e.,
\begin{equation}
\mathscr{D}_F(\bd{A}) = \tfrac12 \| \bd{A} - \Diag \bd{A} \|^2_F.
\end{equation}
This is a general diagonality measure defined for any square matrix
not necessarily Hermitian positive-definite. It goes to zero when
the sum of squared off-diagonal entries goes to zero independently
of the values of the diagonal entries. This measure does not take
positive definiteness into account. Furthermore, while this diagonality
measure satisfies Property~1, it does not satisfy Property~2. 

For a diagonality measure $\mathscr{D}(\cdot)$ to satisfy 
Property~2, it must be such that 
$\mathscr{D}(\bd{A}) = \mathscr{D}(\hat{\bd{A}})$, where for any
positive-definite matrix $\bd{A}$, the matrix $\hat{\bd{A}}$ denotes  
its diagonally scaled matrix given by
\[
\hat{\bd{A}} := (\Diag \bd{A})^{-1/2} \bd{A} (\Diag\bd{A})^{-1/2},
\]
i.e., $\hat{a}_{ij} = \dfrac{a_{ij}}{\sqrt{a_{ii} a_{jj}}}$, $1\le i,j \le n$.
We note that $\hat{a}_{ii} =1$ and $|\hat{a}_{ij} | < 1$ for $1 \le i \ne j \le n$. If $\bd{A}$ is a 
positive-definite covariance matrix then $\hat{\bd{A}}$ is the corresponding 
correlation matrix. Thus, a properly invariant diagonality measure depends 
only on the $n(n-1)/2$ independent entries of the hollow (i.e., with vanishing diagonal) symmetric matrix 
\[
\holmat{\bd{A}} := (\Diag \bd{A})^{-1/2} \bd{A} (\Diag\bd{A})^{-1/2} - \bd{I}
= \hat{\bd{A}} - \bd{I}.
\]

\subsection{Modified Frobenius measure}

Now, using the Frobenius distance between $\hat{\bd{A}}$ and 
$\bd{I}$, we define the modified version of the Frobenius diagonality 
measure
\begin{equation}
\mathscr{D}_F^m(\bd{A}) = \tfrac12 \| (\Diag \bd{A})^{-1/2} \bd{A}(\Diag \bd{A})^{-1/2} - \bd{I} \|^2_F = \tfrac12 \| \hat{\bd{A}} - \bd{I} \|^2_F = \tfrac12 \| \holmat{\bd{A}} \|^2_F,
\end{equation}
which does satisfy Property~2.

\subsection{Riemannian measure}

The Riemannian measure of diagonality of a Hermitian positive-definite
matrix $\bd{A}$ is defined as half the square of the Riemannian distance 
(\ref{eq:RiemDist}) between $\bd{A}$ and $\Diag \bd{A}$, i.e.,
\begin{equation}
\mathscr{D}_R(\bd{A}) := \tfrac12 \| \log((\Diag \bd{A})^{-1/2}\bd{A} (\Diag \bd{A})^{-1/2}) \|^2_F = \tfrac12 \| \log(\hat{\bd{A}}) \|^2_F.
\end{equation}

\subsection{Kullback-Leibler measure}

The \emph{right} Kullback-Leibler measure of diagonality of a symmetric 
positive-definite matrix $\bd{A}$ is defined as the Kullback-Leibler divergence 
\cite{MoakherBatchelor06,Moakher09} between $\bd{A}$ and $\Diag \bd{A}$, i.e.,
\begin{align}
\mathscr{D}_{KL}^r(\bd{A}) & := \tr (\bd{A}^{-1} \Diag \bd{A} - \bd{I}) 
- \log \det (\bd{A}^{-1} \Diag \bd{A}) \nonumber \\ & = 
\tr ((\Diag \bd{A})^{1/2} \bd{A}^{-1} (\Diag \bd{A})^{1/2} - \bd{I}) \nonumber \\  
& \qquad - \log \det ((\Diag \bd{A})^{1/2} \bd{A}^{-1} (\Diag \bd{A})^{1/2}) \nonumber \\
&= \tr \hat{\bd{A}}^{-1} - n + \log \det \hat{\bd{A}}.
\end{align}
We note that the fact $\mathscr{D}_{KL}^r(\bd{A})\ge 0$, with equality if and only if $\bd{A}$ is diagonal, is nothing but the fact that, for a positive-definite matrix $\bd{B}$, $\tr(\bd{B} - \bd{I})$ is an upper bound for $\log \det \bd{B}$, which is attained if and only if $\bd{B}$ is the identity. This follows immediately from the inequality $\log x \le x-1$ which is valid for all $x>0$ and the equality holds if and only if $x=1$.

Similarly, the \emph{left} Kullback-Leibler measure of diagonality of a symmetric positive-definite matrix $\bd{A}$ is defined as the Kullback-Leibler divergence between $\Diag \bd{A}$ and $\bd{A}$, i.e.,
\begin{align}
  \mathscr{D}_{KL}^l(\bd{A}) & := \tr ((\Diag \bd{A})^{-1} \bd{A} - \bd{I})
  - \log \det ((\Diag \bd{A})^{-1} \bd{A}) \nonumber \\ & =
\tr ((\Diag \bd{A})^{-1/2}\bd{A} (\Diag \bd{A})^{-1/2} - \bd{I}) \nonumber \\  
& \qquad - \log \det ((\Diag \bd{A})^{-1/2}) \bd{A} (\Diag \bd{A})^{-1/2})  \nonumber \\
 & = - \log \det \hat{\bd{A}}.
\end{align}
We note that the fact $\mathscr{D}_{KL}^l(\bd{A})\ge 0$, with equality if and 
only if $\bd{A}$ is diagonal, is just a form of Hadamard's 
determinant inequality.

Also, as there are many different ways to symmetrize the Kullback-Leibler
divergence, we may define different corresponding diagonality measures. The simplest one is obtained by taking the arithmetic mean of the right and left ones.
Hence   
\begin{equation}
\mathscr{D}_{KL}^s(\bd{A}) := \tfrac12 (\mathscr{D}_{KL}^r(\bd{A}) + \mathscr{D}_{KL}^l(\bd{A})) = \tfrac12(\tr \hat{\bd{A}}^{-1} - n).
\end{equation}
We see that the symmetrized Kullback-Leibler diagonality measure is simply the inverse-barrier function for the set of Hermitian positive-definite matrices, while the left Kullback-Leibler diagonality measure is simply the logarithmic-barrier function for this set.

\subsection{Log-det $\alpha$-measure}

For $\alpha \in ]-1,1[$, the log-det $\alpha$-measure of diagonality of a 
Hermitian positive-definite matrix $\bd{A}$ is defined as the log-det 
$\alpha$-divergence \cite{Chebbi12} between $\bd{A}$ and $\Diag \bd{A}$, i.e.,
\begin{align} \label{lgdetalph}
\mathscr{D}_{\alpha}(\bd{A}) & := \frac{4}{1-\alpha^2} \log \frac{\det (\tfrac{1-\alpha}{2}\bd{A} + \tfrac{1+\alpha}{2} \Diag \bd{A})}{(\det \bd{A})^{\frac{1-\alpha}{2}} (\det \Diag \bd{A})^{\frac{1+\alpha}{2}}} \nonumber \\
& = \frac{4}{1-\alpha^2} \log \frac{\det (\tfrac{1-\alpha}{2}\hat{\bd{A}} + \tfrac{1+\alpha}{2} \bd{I})}{(\det \hat{\bd{A}})^{\frac{1-\alpha}{2}}}.
\end{align}
The fact that $\mathscr{D}_{\alpha}(\bd{A})\ge 0$, with equality if and only if 
$\bd{A}$ is diagonal, is a consequence of Ky Fan's determinant inequality for 
Hermitian positive-definite matrices $\bd{A}$ and $\bd{B}$:
\[
\det (\lambda \bd{A} + (1-\lambda) \bd{B}) \ge (\det \bd{A})^{\lambda} (\det \bd{B})^{1-\lambda}, \quad \forall \lambda\; \in [0,1].
\]
In the special case $\alpha=0$, we obtain the Bhattacharyya measure of diagonality 
\begin{equation}
\mathscr{D}_B(\bd{A}) := 4 \log \frac{\det \tfrac12(\bd{A} + \Diag \bd{A})}{\sqrt{\det \bd{A} \det \Diag \bd{A}}} = 4 \log \frac{\det \tfrac12(\hat{\bd{A}} + \bd{I})}{\sqrt{\det \hat{\bd{A}}}},
\end{equation}
which is half the square of the Bhattacharyya distance
\cite{Chebbi12} between $\bd{A}$ and $\Diag \bd{A}$.

Using the fact that for an invertible matrix-valued function $\bd{C}(t)$ we have 
\[
\frac{d }{dt}\log \det \bd{C}(t) = \tr\left(\bd{C}^{-1}(t) \frac{d }{dt} \bd{C}(t)\right),
\]
one can verify that the limit of $\mathscr{D}_{\alpha}(\bd{A})$ when $\alpha$ goes to -1 is the right Kullback-Leibler measure and the limit of $\mathscr{D}_{\alpha}(\bd{A})$ when $\alpha$ goes to 1 is the left Kullback-Leibler measure.

\begin{figure}[!htbp]
\caption{Plots of the different diagonality measures for 2-by-2 Hermitian 
positive-definite matrices. Note that, all diagonality measures except the Frobenius one, are defined for $r$ in $[0,1[$.}
\begin{center}
\includegraphics[width=.8\textwidth]{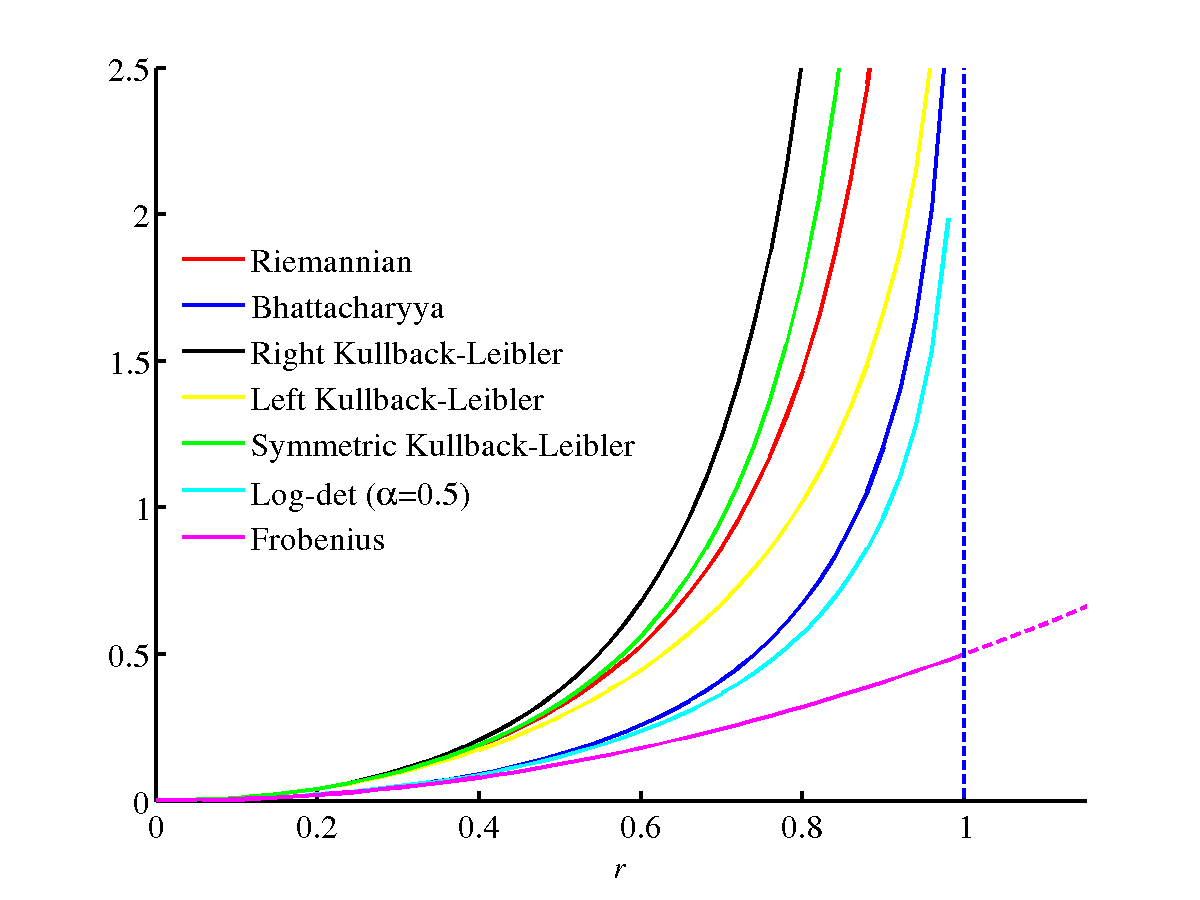}
\end{center}
\label{fig1}
\end{figure}

\begin{remark}
All these diagonality measures, except the modified Frobenius one, 
go to infinity when the matrix $\bd{A}$ approaches the boundary of
the set of Hermitian positive-definite matrices, see Fig.~\ref{fig1} for an illustration of the case of $2 \times 2$ matrices.
\end{remark}

\subsection{Closed-form expressions for the diagonality measures}

\subsubsection{Case of $2 \times 2$ matrices}

In this case, let $z = \frac{a_{12}}{\sqrt{a_{11} a_{22}}}$ and $r= |z|$. Note that $0 \le r < 1$ by positive definiteness of $\bd{A}$. Then,
\[
\hat{\bd{A}} = \begin{bmatrix} 1 & z\\ \bar{z} & 1 \end{bmatrix}, \quad \holmat{\bd{A}} = \begin{bmatrix} 0 & z\\ \bar{z} & 0 \end{bmatrix} \text{ and }
r^2 = \tfrac12 \tr (\holmat{\bd{A}}^2).
\]
Therefore,
\begin{align*}
\mathscr{D}_F^m(\bd{A}) & = r^2, \\
\mathscr{D}_R(\bd{A}) & = \tfrac12 ( \log^2 (1 + r) + \log^2 (1 - r)), \\
\mathscr{D}_{KL}^r(\bd{A}) & = \frac{2 r^2}{1 - r^2} + \log (1 - r^2), \\
\mathscr{D}_{KL}^l(\bd{A}) & = - \log (1 - r^2), \\
\mathscr{D}_{KL}^s(\bd{A}) & = \frac{r^2}{1 - r^2}, \\
\mathscr{D}_{\alpha}(\bd{A}) & = \tfrac{2}{1-\alpha^2}\log (1 - (\tfrac{1-\alpha}{2}r)^2) - \tfrac{1}{(1+\alpha)} \log (1 - r^2), \\
\mathscr{D}_B(\bd{A}) & = 2\log (1 - \tfrac14 r^2) - \log (1 - r^2). 
\end{align*}

Here we give the power series expansions of these functions about $r=0$, i.e., for almost diagonal matrices:
\begin{align*}
\mathscr{D}_F^m(\bd{A}) &= {r}^{2},\\
\mathscr{D}_R(\bd{A}) &= {r}^{2} + \sum_{k=2}^\infty \frac{1}{k} \left( \sum_{j=1}^{2k-1} \tfrac1{j} \right) r^{2k}, \\
\mathscr{D}_{KL}^r(\bd{A}) &= {r}^{2} + \sum_{k=2}^\infty \tfrac{2k-1}{k} r^{2k}, \\
\mathscr{D}_{KL}^l(\bd{A}) &= {r}^{2} + \sum_{k=2}^\infty \tfrac{1}{k} r^{2k}, \\
\mathscr{D}_{KL}^s(\bd{A}) &= {r}^{2} + \sum_{k=2}^\infty r^{2k}, \\
\mathscr{D}_\alpha(\bd{A}) &= {r}^{2} + \sum_{k=2}^\infty \frac1{k} \left(\sum_{j=0}^{k-1} (\tfrac{1-\alpha}{2})^j \right) r^{2k},\\
\mathscr{D}_B(\bd{A}) &= {r}^{2} + \sum_{k=2}^\infty \frac2{k} \left(1 - \tfrac1{2^{k}} \right) r^{2k}.
\end{align*}

\subsubsection{Case of $3 \times 3$ matrices}

In this case, let $a = \frac{a_{12}}{\sqrt{a_{11} a_{22}}}$, $b = \frac{a_{31}}{\sqrt{a_{11} a_{33}}}$ and $c = \frac{a_{23}}{\sqrt{a_{22} a_{33}}}$. Then,
\[
\hat{\bd{A}} = \begin{bmatrix} 1 & a & \bar{b} \\ \bar{a} & 1 & c \\ b & \bar{c} & 1 \end{bmatrix}, \quad \holmat{\bd{A}} = \begin{bmatrix} 0 & a & \bar{b} \\ \bar{a} & 0 & c \\ b & \bar{c} & 0 \end{bmatrix}.
\]
Positive definiteness of $\bd{A}$ implies that $1 - \rho^2 + 2 \delta > 0$ and $|\delta| < 1$ where $\rho$ is the non-negative number given by $\rho^2 = |a|^2 + |b|^2 + |c|^2 = \tfrac12 \tr(\holmat{\bd{A}}^2)$ and $\delta = \Re(a b c) = \tfrac16 \tr(\holmat{\bd{A}}^3)$. 

Using Cardano's trigonometric solution of a cubic equation \cite{Smith61}
the eigenvalues of $\hat{\bd{A}}$ are $1 + s_i$, $i=1,\ldots , 3$ where
\[
s_1 = \frac{2 \rho}{\sqrt{3}} \cos \phi, \quad
s_2 = - \frac{2 \rho}{\sqrt{3}} \cos \left(\frac{\pi}3 + \phi \right), \quad
s_3 = - \frac{2 \rho}{\sqrt{3}} \cos \left(\frac{\pi}3 - \phi \right),
\]
with $\phi = \tfrac13 \cos^{-1} \dfrac{3\sqrt{3} \delta}{\rho^3}$.

Then, the diagonality measures are given explicitly by
\begin{align*}
\mathscr{D}_F^m(\bd{A}) & = \rho^2, \\
\mathscr{D}_R(\bd{A}) & = \tfrac12 ( \log^2 (1 + s_1) + \log^2 (1 + s_2) + \log^2(1 + s_3)), \\
\mathscr{D}_{KL}^r(\bd{A}) & = \frac{2 (\rho^2 - 3\delta)}{1 - \rho^2 + 2\delta} + \log (1 - \rho^2 + 2\delta), \\
\mathscr{D}_{KL}^l(\bd{A}) & = - \log (1 - \rho^2 + 2\delta), \\
\mathscr{D}_{KL}^s(\bd{A}) & = \frac{\rho^2 - 3\delta}{1 - \rho^2 + 2\delta}, \\
\mathscr{D}_{\alpha}(\bd{A}) & = \tfrac{2}{1 - \alpha^2}\log (1 - (\tfrac{1-\alpha}{2}\rho)^2 + 2 (\tfrac{1-\alpha}2)^3 \delta) - \tfrac{1}{1-\alpha} \log (1 - \rho^2 + 2\delta), \\
\mathscr{D}_B(\bd{A}) & = 2\log (1 - \tfrac14 \rho^2 +\tfrac14 \delta) - \log (1 - \rho^2 + 2\delta). 
\end{align*}
We remark that all measures depend only on the two parameters $\rho$ and $\delta$. Furthermore, it is interesting to note that if $\delta$ is equal to zero, i.e., when at least one of the three off-diagonal entries $a$, $b$ or $c$ is zero, then all these diagonality measures have the same functional expressions as in the $2 \times 2$ case (with $r$ replaced by $\rho$).

\subsubsection{The general case of $n \times n$ matrices}

In the general case of $n \times n$ matrices, as we shall see, it is possible to have series expansions of the diagonality measures in terms of (only) the $n-1$ invariants, $\tr (\holmat{\bd{A}}^i), \ i=2, \ldots, n$, of the hollow matrix $\holmat{\bd{A}}$. Indeed, as the spectral radius of $\holmat{\bd{A}}$ is less than one, we have
$\log (\bd{I} + \holmat{\bd{A}}) = \sum_{k=1}^\infty \frac{(-1)^{k-1}}{k}  \holmat{\bd{A}}^k$, and $(\bd{I} + \holmat{\bd{A}})^{-1} = \sum_{k=0}^\infty (-1)^{k} \holmat{\bd{A}}^k$. Therefore, as $\holmat{\bd{A}}$ is traceless, we obtain
\begin{align*}
\mathscr{D}_F^m (\bd{A}) & = \tfrac12 \tr (\holmat{\bd{A}}^2), \\
\mathscr{D}_R (\bd{A}) & = \tfrac12 \tr (\holmat{\bd{A}}^2) + 
\sum_{k=3}^\infty  \frac{(-1)^{k}}{k} \left(\sum_{j=0}^{k-2} \frac{1}{j+1} \right) \tr (\holmat{\bd{A}}^{k}), \\
\mathscr{D}_{KL}^r (\bd{A}) & = \tfrac12 \tr (\holmat{\bd{A}}^2) + \sum_{k=3}^\infty (-1)^{k} \frac{k-1}{k} \tr (\holmat{\bd{A}}^{k}), \\ 
\mathscr{D}_{KL}^l (\bd{A}) & = \tfrac12 \tr (\holmat{\bd{A}}^2) + \sum_{k=3}^\infty \frac{(-1)^{k}}{k} \tr (\holmat{\bd{A}}^{k}), \\ 
\mathscr{D}_{KL}^s (\bd{A}) & = \tfrac12 \tr (\holmat{\bd{A}}^2) + \sum_{k=3}^\infty \frac{(-1)^{k}}2 \tr (\holmat{\bd{A}}^{k}), \\ 
\mathscr{D}_{\alpha} (\bd{A}) & = \tfrac12 \tr (\holmat{\bd{A}}^2) + \sum_{k=3}^\infty \frac{(-1)^{k}}{k} \left( \sum_{j=0}^{k-2}\left(\frac{1-\alpha}{2}\right)^{j}\right) \tr (\holmat{\bd{A}}^{k}), \\ 
\mathscr{D}_{B} (\bd{A}) & = \tfrac12 \tr (\holmat{\bd{A}}^2) + \sum_{k=3}^\infty \frac{2(-1)^{k}}{k} \left(1 - \frac{1}{2^{k-1}}\right) \tr (\holmat{\bd{A}}^{k}).
\end{align*}
Once again, all these measures agree to second order. Now we show that they depend only on the $n-1$ invariants, $T_i := \tr (\holmat{\bd{A}}^i), \ i=2, \ldots, n$, of the hollow matrix $\holmat{\bd{A}}$. Indeed, using the Cayley-Hamilton theorem and since $\holmat{\bd{A}}$ is a traceless matrix, we have
\[
\holmat{\bd{A}}^n = (-1)^n \sum_{i=2}^n b_{i} \holmat{\bd{A}}^{n-i},
\]
where the coefficients $b_i$ are given by the recursive formula \cite{Horst35}
\[
b_i = \frac1{i} \left( \sum_{j=2}^{i-2} b_{i-j} T_j + (-1)^n T_i \right), \quad i=2, \ldots, n.
\]
So any power $\holmat{\bd{A}}^p$ of $\holmat{\bd{A}}$ with $p > n$ can be expressed in terms of powers $\holmat{\bd{A}}^q$ of $\holmat{\bd{A}}$ of order $2 \le q \le n$. The result then follows by taking the trace.

\section{True diagonality measures} \label{sec:tdiag}

The ``true" diagonality measure is defined as the distance (divergence) between a given positive-definite matrix and its closest, with respect to the distance (divergence), diagonal one.

The closest positive-definite diagonal matrix with respect to a distance $d(\cdot, \cdot)$ is defined as
\[
\argmin{\bd{X} \in \mathscr{D}_n^{++}} d(\bd{A}, \bd{X}).
\]
Similarly, the \emph{right} closest positive-definite diagonal matrix with respect to a divergence $D(\cdot, \cdot)$ is defined as
\[
\argmin{\bd{X} \in \mathscr{D}_n^{++}} D(\bd{A}, \bd{X}),
\]
and the \emph{left} closest positive-definite diagonal matrix with respect to a divergence $D(\cdot, \cdot)$ is defined as
\[
\argmin{\bd{X} \in \mathscr{D}_n^{++}} D(\bd{X}, \bd{A}).
\]

The closest, with respect to the Frobenius distance, diagonal matrix to $\bd{A}$ is $\bd{D}_F = \Diag(\bd{A})$, whereas the closest, with respect to the Riemannian distance, is $\bd{D}_R$ the unique positive-definite diagonal solution to 
\begin{equation} \label{eq:closestR}
\Diag( \log(\bd{A}^{-1} \bd{X})) = \bd{0}.
\end{equation}
Equation (\ref{eq:closestR}) can be solved numerically by using the steepest descent method for finding the minimum of the convex function $\bd{X} \mapsto d_{\mathscr{H}^{++}_n}(\bd{A}, \bd{X})$ defined on the set of positive-definite diagonal matrices. Starting from an initial guess $\bd{X}^{(0)}$, we compute the iterates:
\[
\bd{X}^{(k+1)} = \bd{X}^{(k)} - r \bd{X}^{(k)}\Diag( \log(\bd{A}^{-1} \bd{X}^{(k)})),
\]
where $r$ is a conveniently chosen stepsize.

The right closest, with respect to the Kullback-Leibler divergence, diagonal matrix to $\bd{A}$ is $\bd{D}_{KL}^r = \Diag(\bd{A})$, and the left closest, with respect to the Kullback-Leibler divergence, diagonal matrix to $\bd{A}$ is $\bd{D}_{KL}^l = (\Diag(\bd{A}^{-1})^{-1})$.
For the symmetrized Kullback-Leibler divergence, the closest diagonal matrix to $\bd{A}$ is $\bd{D}_{KL}^s = \Diag(\bd{A})^ {1/2}(\Diag(\bd{A}^{-1}))^{-1/2}$.

The right closest, with respect to the log-det $\alpha$-divergence, diagonal matrix to $\bd{A}$ is $\bd{D}_{LD}^\alpha$, the unique positive-definite diagonal matrix solution of 
\begin{equation} \label{eq:closestalpha}
\Diag\left( \left(\frac{1-\alpha}2 \bd{A} + \frac{1+\alpha}2 \bd{X} \right)^{-1} \right) = \bd{X}^{-1}.
\end{equation}
So, for the Bhattacharyya divergence, the closest diagonal matrix to $\bd{A}$ is $\bd{D}_B$ the unique positive-definite diagonal matrix solution of 
\[
2 \Diag\left( (\bd{A} + \bd{X})^{-1} \right) = \bd{X}^{-1}.
\]
Equation (\ref{eq:closestalpha}) can be solved numerically by using the steepest descent method for finding the minimum of the convex function $\bd{X} \mapsto D^{\alpha}_{LD}(A, \bd{X})$ defined on the set of positive-definite diagonal matrices.

We note that $\Diag \bd{A}$ is the closest diagonal matrix to $\bd{A}$ only for the Frobenius distance and right Kullback-Leibler divergence.

\section{Application to the approximate joint diagonalization problem} \label{sec:jad}

In Congedo et al. \cite{congedo15} it was shown that the Riemannian metric-based geometric mean \cite{Moakher05,bhatia07,bhatia06} of a set of Hermitian positive-definite matrices $\bd{M}_k$, $k=1, \ldots, K$, can be approximated numerically by the approximate joint diagonalizer of the set as
\begin{equation} \label{gmean}
\bd{G} \approx \bd{C}^{-1} \exp \left( \frac1{k} \sum_{k=1}^K \log \bd{D}_k \right) \bd{C}^{-H},
\end{equation}
where $\exp \left( \frac1{k} \sum_{k=1}^K \log \bd{D}_k \right)$ is the Log-Euclidean mean \cite{arsigny2007} of the transformed set $\bd{D}_k = \bd{C} \bd{M}_k \bd{C}^H$, which, thanks to the congruence invariance of the geometric mean, coincides with the transformed geometric mean $\bd{C} \bd{G} \bd{C}^H$ when the transformed matrices all pair-wise commute. More generally, using the same argument, we may approximate all power means of order $p\in [-1, 1]$ (see \cite{lim2012}) by the approximate joint diagonalizer as
\begin{equation} \label{pmean}
\bd{G}_p \approx \bd{C}^{-1} \left( \frac1{k} \sum_{k=1}^K \bd{D}_k^p \right)^{1/p} \bd{C}^{-H},
\end{equation}
where, $\left( \frac1{k} \sum_{k=1}^K \bd{D}_k^p\right)^{1/p}$, is the power mean of transformed matrices. Approximation (\ref{gmean}) is the limit of (\ref{pmean}) when $p \to 0$ and, of course, for $p=1$ and $p=-1$ we have the arithmetic mean and harmonic mean, respectively. From a computational point of view the approximate joint diagonalizer approach to compute means may take advantage of the quadratic convergence displayed by some approximate joint diagonalizer algorithms \cite{congedo15}. In \cite{congedo15} it has been shown that the properties of approximation (\ref{gmean}) inherit directly from the properties of the cost function used to find the joint diagonalizer $\bd{C}$; using Pham's cost function the approximation verifies the determinant identity, the joint homogeneity, invariance by congruence transformation, but not the self-duality. An approximate joint diagonalization criterion satisfies the self-duality property whenever if $\bd{C}$ is an approximate joint diagonalizer of the set $\bd{M}_1, \ldots, \bd{M}_K$, then $\bd{C}^{-H}$ is an approximate joint diagonalizer of the set $\bd{M}_1^{-1}, \ldots, \bd{M}_K^{-1}$. Pham's criterion does not satisfy self-duality because it is asymmetric. Using a symmetric criterion the self-duality of the resulting mean would be satisfied as well, thus we would obtain approximations of means with all desirable properties. This extends to all power means. Motivated by this and by recent advances in $\alpha$-divergence means resulting from the use of the log-barrier function \cite{Chebbi12,palfia2016,sra2015}, in this article we study approximate joint diagonalization criteria from the perspective of information geometry. While the Riemannian criterion is difficult to handle, we give the gradient and Hessian for a new approximate joint diagonalization cost function based on the $\alpha$-divergence. For $\alpha=0$ the resulting approximate joint diagonalization criterion is symmetric, as desired for estimating means of Hermitian positive-definite matrices based on their approximate joint diagonalizer.

The most commonly used cost function in the problem of joint diagonalization of a set of $K$ covariance matrices $\{ \bd{M}_k \in \mathscr{H}^{++}_n, \ k=1, \ldots, K \}$, is Pham's cost function given by \cite{Flury86, Flury84, pham01}
\begin{equation} \label{eq:J}
\hspace*{-.2cm}
\mathcal{J}(\bd{C}; \{ \bd{M}_k \}) \!=\! \sum_{k=1}^{K} \! \beta_k \! \left[
\log\left(\det(\Diag(\bd{C}\bd{M}_k\bd{C}^H))\right) \!-\!
\log\left(\det(\bd{C}\bd{M}_k\bd{C}^H)\right) \right],
\end{equation}
where $\beta_k$ are positive weights. We remark that the above cost function is nothing but the sum of the weighted left Kullback-Leibler diagonality measures of $\bd{C}\bd{M}_k\bd{C}^H$, $k=1, \ldots, K$.

We here propose a one-parameter family of cost functions that measure the degree of joint diagonalization of a set of Hermitian positive-definite matrices which is more general than (\ref{eq:J}) and which includes it as a special case.
The family of cost functions $\mathcal{J}_{\alpha}$, parameterized by $\alpha \in [-1, 1]$, that we propose here is defined as the sum of the log-det $\alpha$-diagonality measures of the matrices $\bd{C}\bd{M}_k\bd{C}^H$, i.e.,
\begin{equation} \label{costfunction}
\mathcal{J}^{(\alpha)}(\bd{C}; \{ \bd{M}_k \}) = \sum_{k=1}^{K} \tilde{\mathcal{J}}^{(\alpha)}_{\bd{M}_k}(\bd{C}),
\end{equation}
where 
\begin{equation} \label{logdt}
\tilde{\mathcal{J}}^{(\alpha)}_{\bd{M}}(\bd{C}) = \mathscr{D}_{\alpha}(\bd{C} \bd{M} \bd{C}^H),
\end{equation}
is the log-det $\alpha$-diagonality measure of the matrix $\bd{C} \bd{M} \bd{C}^H$. It should be noted that the log-determinant $\alpha$-measure of diagonality covers, through the parameter $\alpha$, all (except the modified Frobenius and Riemannian) diagonality measures discussed in Section~\ref{sec:diag}. We remark that the numerical implementation of an approximate joint diagonalizer based on the Riemannian diagonality measure can be quite complicated. However, the Riemannian diagonality measure can be well approximated by the symmetrized Kullback-Leibler diagonality measure.

The problem of joint diagonalization is then formulated as the following minimization problem
\[
\min_{\bd{C} \in {\rm GL}_n(\mathbb{C})} \mathcal{J}^{(\alpha)}(\bd{C}; \{ \bd{M}_k \}),
\]
where $\alpha$ is a real parameter in $[-1,1]$ and $\{ \bd{M}_k \}$ is a given set of $K$ Hermitian positive-definite matrices. To numerically solve this minimization problem, we will use a modified Newton method \cite{Joho08}. For this method we need to obtain explicit expressions for the gradient and Hessian of the cost function.

\subsection{Gradient and Hessian of the cost function}

In order to represent the gradient and the Hessian of the cost function (\ref{costfunction}) in a compact form, we use the matrix-form representation of the second-order Taylor series expansion as described by Manton in \cite{manton2002optimization} and which we recall here. For a twice differentiable function $\mathcal{J} : \mathcal{M}_{n}(\mathbb{C}) \rightarrow \mathbb{R}$, the second-order Taylor approximation of $\mathcal{J}(\cdot)$ at $\bd{C}$ reads
\begin{align} \label{taylor}
\mathcal{J}(\bd{C} + t \bd{Z}) & = \mathcal{J}(\bd{C}) + t \mathfrak{Re} \left\{ \tr(\bd{Z}^H \bd{G}({\bd{C}}))\right\} + \frac{t^2}{2}\vec{(\bd{Z})}^H \bd{H}({\bd{C}}) \vec{(\bd{Z})} \nonumber \\
  & \quad + \frac{t^2}{2}\mathfrak{Re}\left\{\vec{(\bd{Z})}^T\bd{S}({\bd{C}}) \vec{(\bd{Z})}\right\} + O(t^3),
\end{align}
where $\bd{C}$, $\bd{Z} \in \mathcal{M}_{n}(\mathbb{C})$, $\bd{G}({\bd{C}}) \in \mathcal{M}_{n}(\mathbb{C})$ is the gradient of $\mathcal{J}(\cdot)$ evaluated at $\bd{C}$, and $\bd{H}({\bd{C}})$, $\bd{S}({\bd{C}}) \in \mathcal{M}_{n^2}(\mathbb{C})$ are the Hessians of $\mathcal{J}$ evaluated at $\bd{C}$. To ensure uniqueness of this representation, $\bd{H}({\bd{C}})$ is required to be Hermitian and $\bd{S}({\bd{C}})$ is required to be symmetric, i.e., $\bd{H}({\bd{C}})^H = \bd{H}({\bd{C}})$ and $\bd{S}({\bd{C}})^T = \bd{S}({\bd{C}})$.

To simplify the derivation of the gradient and Hessian of $\mathcal{J}^{(\alpha)}(\; \cdot \; ; \{ \bd{M}_k \})$, it is important to note that $\tilde{\mathcal{J}}^{(\alpha)}_{\bd{M}}(\bd{C})$, defined in (\ref{logdt}), can be conveniently  written as 
\begin{equation} \label{digms}
\tilde{\mathcal{J}}^{(\alpha)}_{\bd{M}}(\bd{C}) = \frac{4}{1-\alpha^2} \mathcal{J}^{(\alpha)}_{\bd{M}}(\bd{C}) - \frac{2}{1+\alpha} \mathcal{J}^{(-1)}_{\bd{M}}(\bd{C}) - \frac{2}{1-\alpha} \mathcal{J}^{(1)}_{\bd{M}}(\bd{C}),
\end{equation}
where
\begin{equation} \label{e1}
\mathcal{J}^{(\alpha)}_{\bd{M}}(\bd{C}) := \log \det\left(\frac{1-\alpha}{2} \bd{C} \bd{M}\bd{C}^H + \frac{1+\alpha}{2} \Diag(\bd{C} \bd{M}\bd{C}^H) \right). 
\end{equation}

To obtain the gradient and Hessian of $\mathcal{J}^{(\alpha)}(\bd{C} ; \{ \bd{M}_k \})$ it is clear from the expression given in (\ref{digms}) that it suffices to derive the expressions of the gradient and Hessian of $\mathcal{J}^{(\alpha)}_{\bd{M}}(\cdot)$. In fact, the gradient and Hessian of the function $\mathcal{J}^{\alpha}(\; \cdot \; ; \{\bd{M}_k \})$ are given by
\begin{align*}
\bd{G}^{(\alpha)}(\bd{C}; \{\bd{M}_k \}) &= \sum_{k=1}^{K}\widetilde{\bd{G}}_{\bd{M}_k}^{(\alpha)}(\bd{C}), \\
\bd{H}^{(\alpha)}(\bd{C}; \{\bd{M}_k \}) &= \sum_{k=1}^{K}\widetilde{\bd{H}}_{\bd{M}_k}^{(\alpha)}(\bd{C}), \\
\bd{S}^{(\alpha)}(\bd{C}; \{\bd{M}_k \}) &= \sum_{k=1}^{K}\widetilde{\bd{S}}_{\bd{M}_k}^{(\alpha)}(\bd{C}),
\end{align*}
where $\widetilde{\bd{G}}_{\bd{M}}^{(\alpha)}(\bd{C})$, $\widetilde{\bd{H}}_{\bd{M}}^{(\alpha)}(\bd{C})$ and $\widetilde{\bd{S}}_{\bd{M}}^{(\alpha)}(\bd{C})$ are the gradient and Hessian of $\widetilde{\mathcal{J}}^{(\alpha)}_{\bd{M}}(\cdot)$ and whose expressions are given in \ref{appenB}.

Once we have derived the gradient and Hessian of our cost function, we now describe a modified Newton.
The update at the $\ell$-th iteration is given by
\[
\bd{C}^{(\ell+1)} = \bd{C}^{(\ell)} + \mu_{\ell} \bd{W}^{(\ell)},
\]
where $\bd{W}^{(\ell)}$ is the search direction and $\mu_{\ell}$ is the step size at iteration $\ell$. For the Newton method, and since our cost function is not quadratic, we use a modified Newton iteration which includes an Armijo line search. If we are close enough to the minimum the modified Newton update will approach the pure Newton iteration. We note that, as the cost function is not quadratic, the Hessian (whose inverse is needed in the line search) of the cost function can have negative eigenvalues. By adding an appropriate multiple of the identity, the Hessian and hence its inverse can be made positive definite. For implementation issues and further details the reader is referred to \cite{Joho08}.

\subsection{Numerical experiments} 

In this section we perform a comparative study of the convergence and performance of the approximate joint diagonalizer algorithm, which is based on the log-det $\alpha$-diagonality measure and implemented using the Newton method, with two state-of-the-art competing approximate joint diagonalizer algorithms.
For the approximate joint diagonalization of the set $\{ \bd{R}_k\}_{k=1}^K$ of symmetric positive-definite matrices we use three types of algorithms, the {\tt Jadiag} algorithm \cite{pham01}, which is based on successive transformation operating on a pair of rows and columns, similar to the Jacobi method (but the transformation matrix is not restricted to be orthogonal), the {\tt Uwedge} algorithm \cite{tichavsky09}, which is a low complexity approximate joint diagonalizer algorithm that incorporates nontrivial block-diagonal weight matrices into a weighted least squares approximate joint diagonalization criterion, and {\tt LD-Newton}, which is based on the log-det $\alpha$-diagonality measure and implemented using the Newton method. It has been shown that the {\tt Jadiag} and the {\tt Uwedge} algorithms have good convergence properties and fast implementation, which gives the comparison process a significant meaning for the evaluation of our method. For the three algorithms, all iterations are started from the identity  matrix $\bd{C}_{(0)} = \bd{I}_ {N\times N}$ as the initial joint diagonalizer of the set $\{ \bd{R}_k\}_{k=1}^K$.

In many engineering applications, the matrix condition number of the symmetric positive-definite matrix summarizing the data (observations, recordings, etc.) tends to be positively correlated with the number of sensors. Moreover, the dispersion of the matrices is proportional to the noise level. These properties can be reproduced by the following generating model for the $N \times N$ input data matrices $\{ \bd{R}_k\}_{k=1}^{K}$ \cite{congedo16}:
\begin{equation} \label{eq:data}
\bd{R}_k = \bd{A} \bd{D}_k \bd{A}^T + \nu (\bd{V}_k \bd{E}_k \bd{V}_k^T + \mu \bd{I}),
\end{equation}
where, $\bd{A} \bd{D}_k \bd{A}^T$ is the signal part, $\nu \bd{V}_k\bd{E}_k\bd{V}_k^T$ is the structured noise part and $\nu \mu \bd{I}$ (we take $\mu$ such that $\nu \mu = 10^{-6}$) is the uncorrelated noise part. Here, $\bd{A}$ is a matrix with elements drawn at random at each simulation from a uniform distribution in $[-1,1]$ and then normalized so as to have columns with unit norm; $\bd{D}_k$ are $K$ diagonal matrices with diagonal elements $d_{k,n}$ randomly drawn at each simulation from a squared Gaussian distribution with expectation $2^{-n}$, where $n\in \{ 1, \dots, N \}$ is the index of the $N$ diagonal elements; $\bd{V}_k$ are matrices generated as $\bd{A}$ above; $\bd{E}_k$ are matrices generated as $\bd{D}_k$ above and $\nu$ is a constant controlling the SNR of the generated matrices through 
\[
\text{SNR} = \dfrac{\left[\tr\left(\sum_k \bd{A} \bd{D}_k \bd{A}^T \right)\right]}{\nu \left[ \tr\left(\sum_k (\bd{V}_k \bd{E}_k \bd{V}_k^T + \mu \bd{I}) \right)\right] }.
\]

In Table~\ref{tab:cpu} we report the CPU time required for each algorithm (implemented in Matlab R2015a running on an Intel Core i5 processor with 4GB RAM on Windows 7 Professional) to converge. For the numerical convergence, we have used the following stopping criterion 
\[
\frac1{N} \| \bd{C}^{-1}_{(k)} \bd{C}_{(k-1)} - \bd{I} \|^2_F \le \epsilon,
\]
where $\bd{C}_{(k-1)}$ and $\bd{C}_{(k)}$ are two successive iterates. The value of $\epsilon$ is chosen to be very small (1e−15). We note that this criterion does not depend on the size nor on the norm of the matrices. 

\begin{center}
\begin{table}[!htbp]
\centering
\caption{The CPU time (in seconds) needed for convergence for each of the {\tt Jadiag}, {\tt Uwedge} and {\tt LD-Newton} (with different values of the parameter $\alpha$) algorithms in three simulations with different noise levels.}
\label{tab:cpu}
\begin{tabular}{|l|l|l|l|l|}
\hline
{\bf Method} & {\tt Jadiag}            & {\tt Uwedge}      & \multicolumn{2}{c|}{\tt LD-Newton} \\ \hline
\multirow{3}{*}{SNR $=100$} & \multirow{3}{*}{0.9360} & \multirow{3}{*}{0.1560} & $\alpha=-0.75$         & 1.6320        \\ \cline{4-5} 
                        &&                         & $\alpha=0$             & 1.2048        \\ \cline{4-5} 
                        &&                         & $\alpha=0.75$          & 1.6025        \\ \hline
\multirow{3}{*}{SNR $=10$} & \multirow{3}{*}{2.4804} & \multirow{3}{*}{1.3276} & $\alpha=-0.75$         & 3.8510        \\ \cline{4-5} 
                        &&                         & $\alpha=0$             & 3.0351        \\ \cline{4-5} 
&&                         & $\alpha=0.75$          & 3.6124        \\ \hline
\multirow{3}{*}{SNR $=1$} & \multirow{3}{*}{3.7468} & \multirow{3}{*}{1.9366} & $\alpha=-0.75$         & 5.1097        \\ \cline{4-5} 
                        &&                         & $\alpha=0$             & 4.1209        \\ \cline{4-5} 
                        &&                         & $\alpha=0.75$          & 4.7012        \\ \hline
\end{tabular}
\end{table}
\end{center}

Since the cost functions of the three algorithms are different, to compare their performance it is meaningful to use the normalized Amari-Moreau performance index \cite{macchi93}, which measures the closeness between the estimated joint diagonalizer $\bd{C}^{e}$ (the estimated unmixing matrix) and the true unmixing matrix $\bd{A}^{-1}$, defined by
\[
\text{PI}(\bd{M}) = \frac{1}{2p(p-1)} \!\left(\sum_{i=1}^p \!\left( \sum_{j=1}^p \frac{|m_{ij}|}{\max_k|m_{ik}|} - 1 \!\right) \!+\! \sum_{j=1}^p \left( \sum_{i=1}^p \frac{|m_{ij}|}{\max_k |m_{kj}|} - 1 \!\right) \!\right),
\]
where $m_{ij}$ are the elements of the matrix $\bd{M} = \bd{C}^{eT} \bd{A}$ and $p$ is the number of sources.

The Amari-Moreau performance index is commonly used to estimate the efficiency of (joint) blind source separation algorithms when both the mixing and unmixing matrices are known. This index is invariant under the scaling and permutation ambiguities. Its values are between 0 and 1 and the lower it is the better performance we have.

\begin{center}
\begin{figure}[!htbp]
\caption{Plots of the Amari-Moreau index vs the iteration number using {\tt Jadiag}, {\tt Uwedge} and {\tt LD-Newton} (with different values of the parameter $\alpha$) algorithms. Left: SNR = 100, center: SNR = 10 and right: SNR = 1. \label{fig:ami-iter}}
\centering
	\subfloat {\includegraphics[width=.33\textwidth]{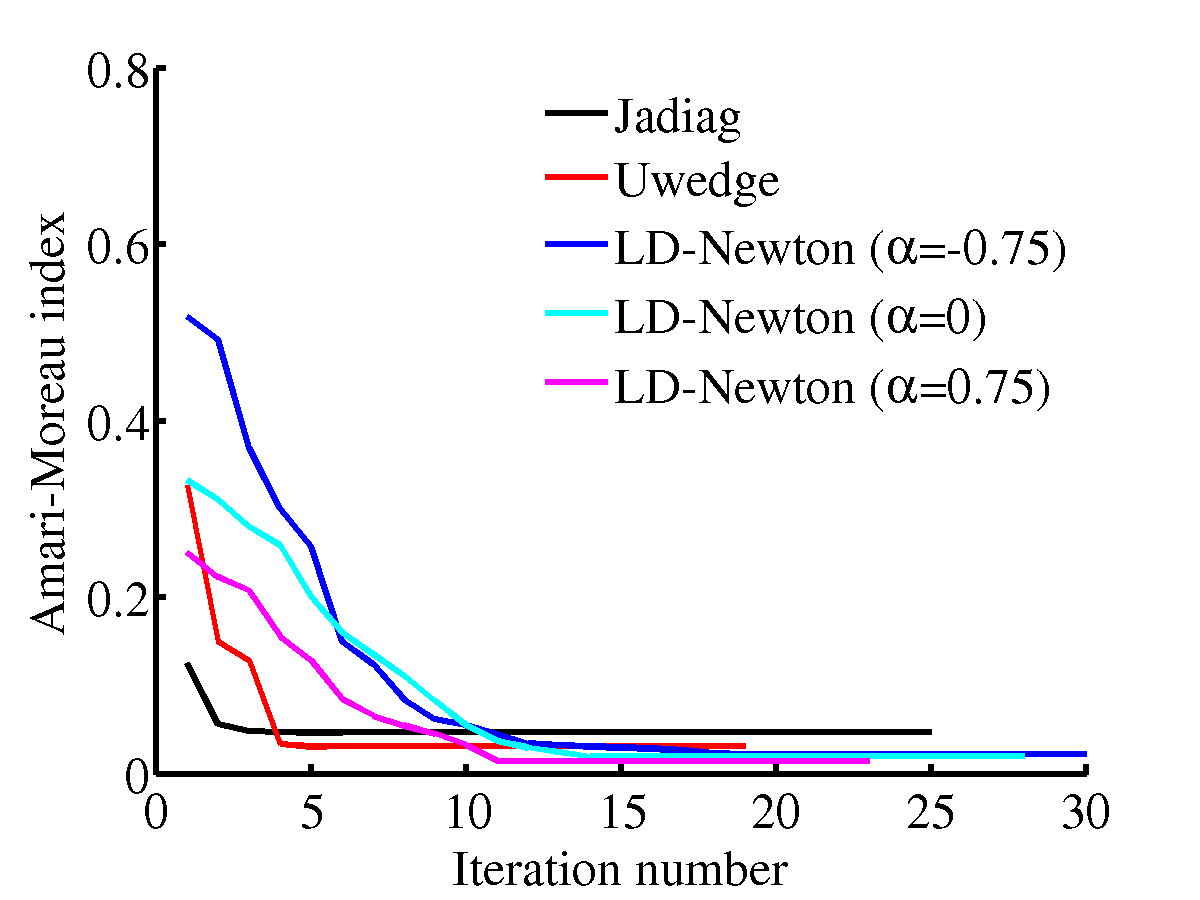}}
 \hfill
	\subfloat  {\includegraphics[width=.33\textwidth]{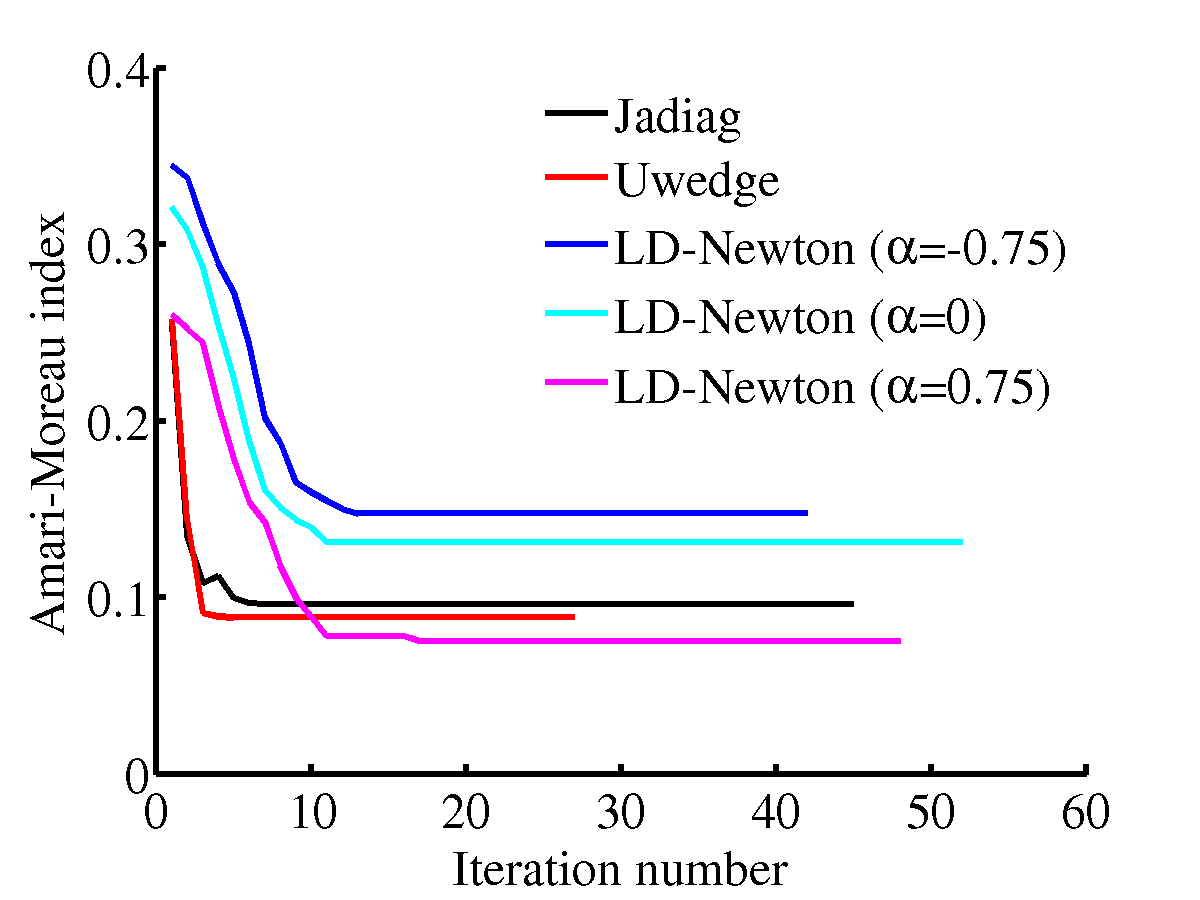}}
 \hfill
	\subfloat  {\includegraphics[width=.33\textwidth]{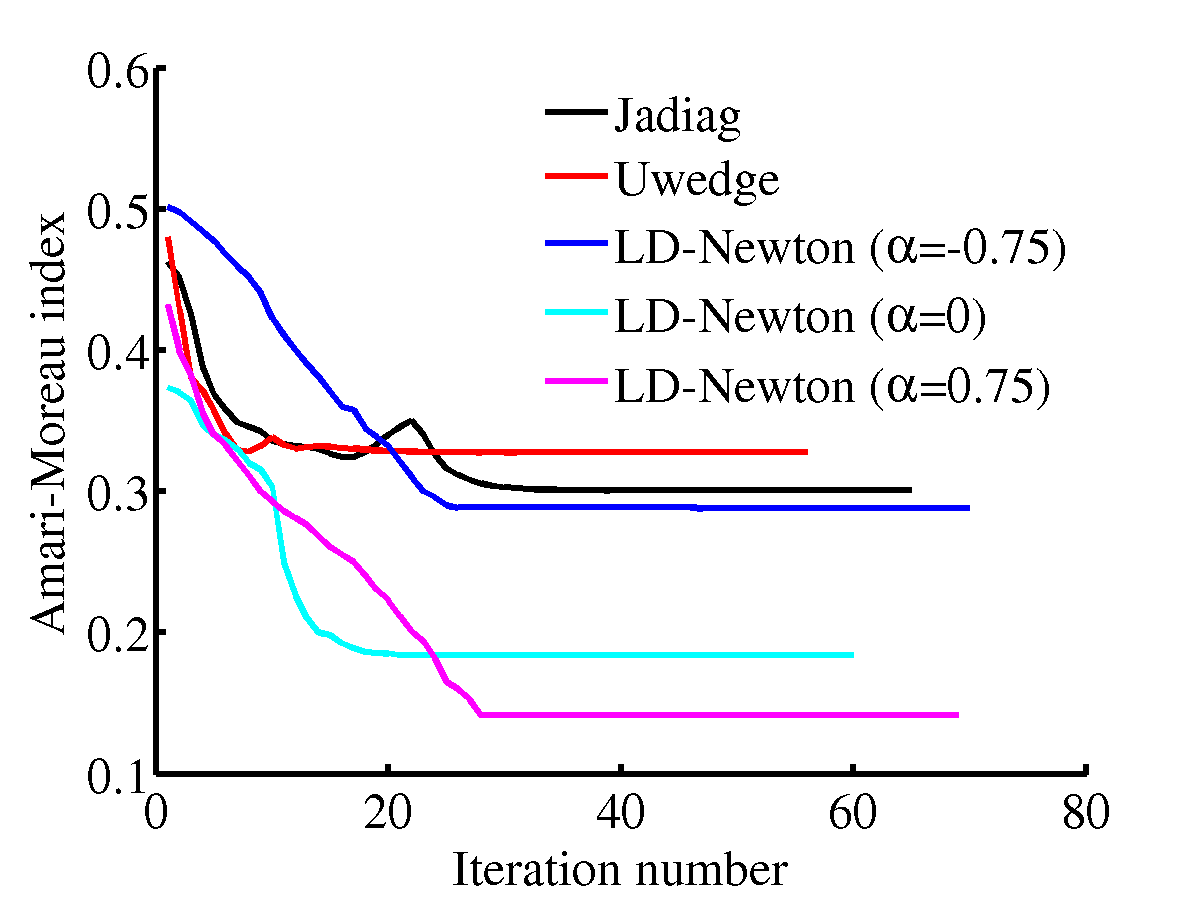}}
\end{figure}
\end{center}

In Fig.~\ref{fig:ami-iter} we show an example of convergence plots for the three algorithms, reporting the evolution of the Amari-Moreau index with respect to the iteration number in three simulations. We performed experiments with data matrices are of size $N=10$ and different SNR taking values in $\{ 1, 10, 100 \}$. For the {\tt LD-Newton} algorithm, the parameter $\alpha$ takes the values $\{ -0.75, 0, 0.75\}$.

\begin{center}
\begin{figure}[!htbp]
\caption{Plots of the Amari-Moreau index for $K=100$ simulations using the {\tt Jadiag}, {\tt Uwedge} and {\tt LD-Newton} (with different values of the parameter $\alpha$) algorithms for two data sizes ($N=20$ left, $N=50$ right). \label{fig:ami-K}}
\centering
	\subfloat{\includegraphics[width=.4\textwidth]{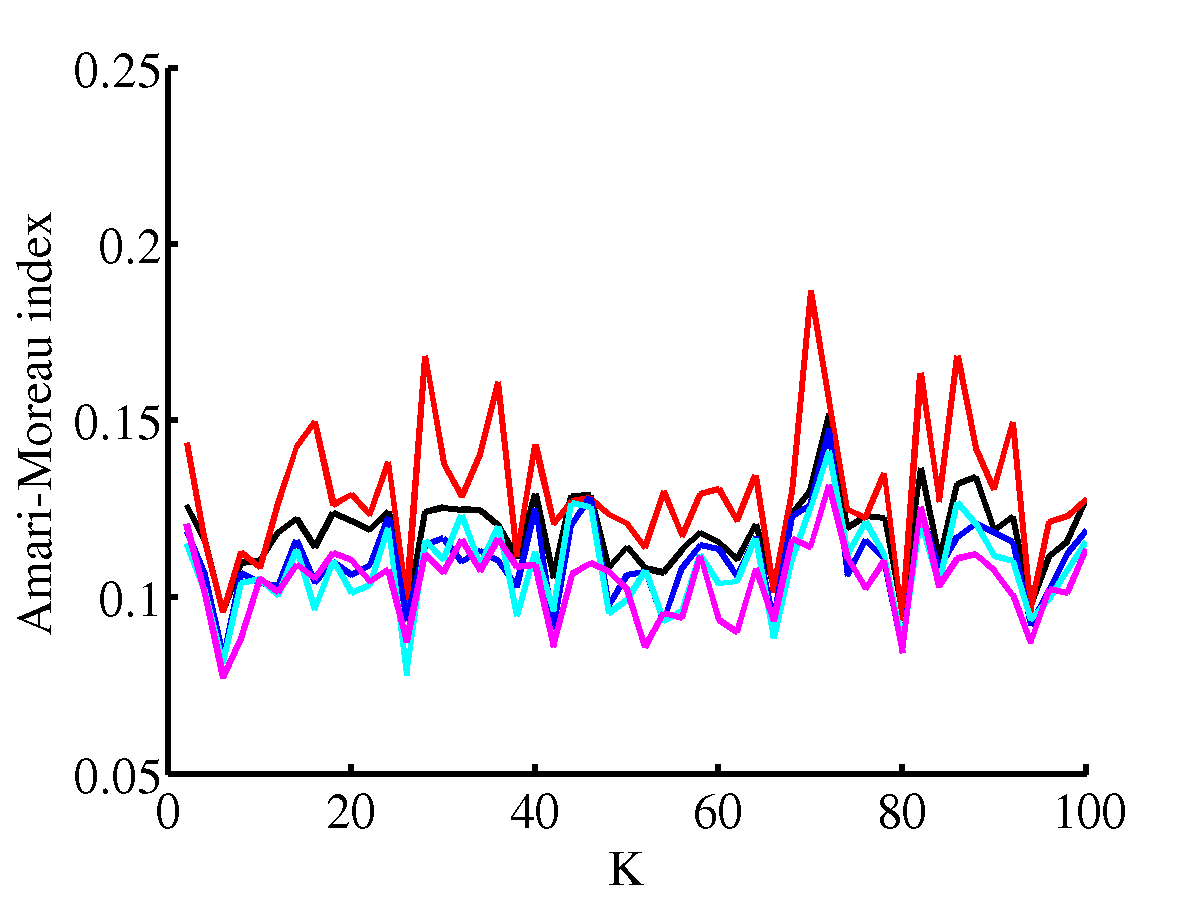}}
        \
        \subfloat{\includegraphics[width=.15\textwidth]{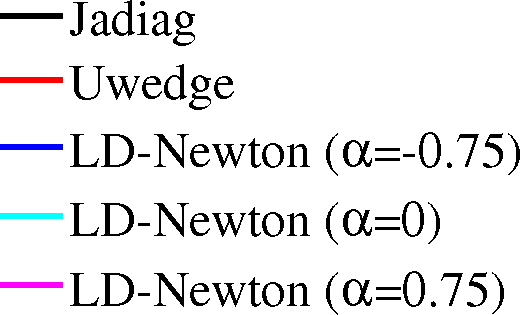}}
        \
	\subfloat{\includegraphics[width=.4\textwidth]{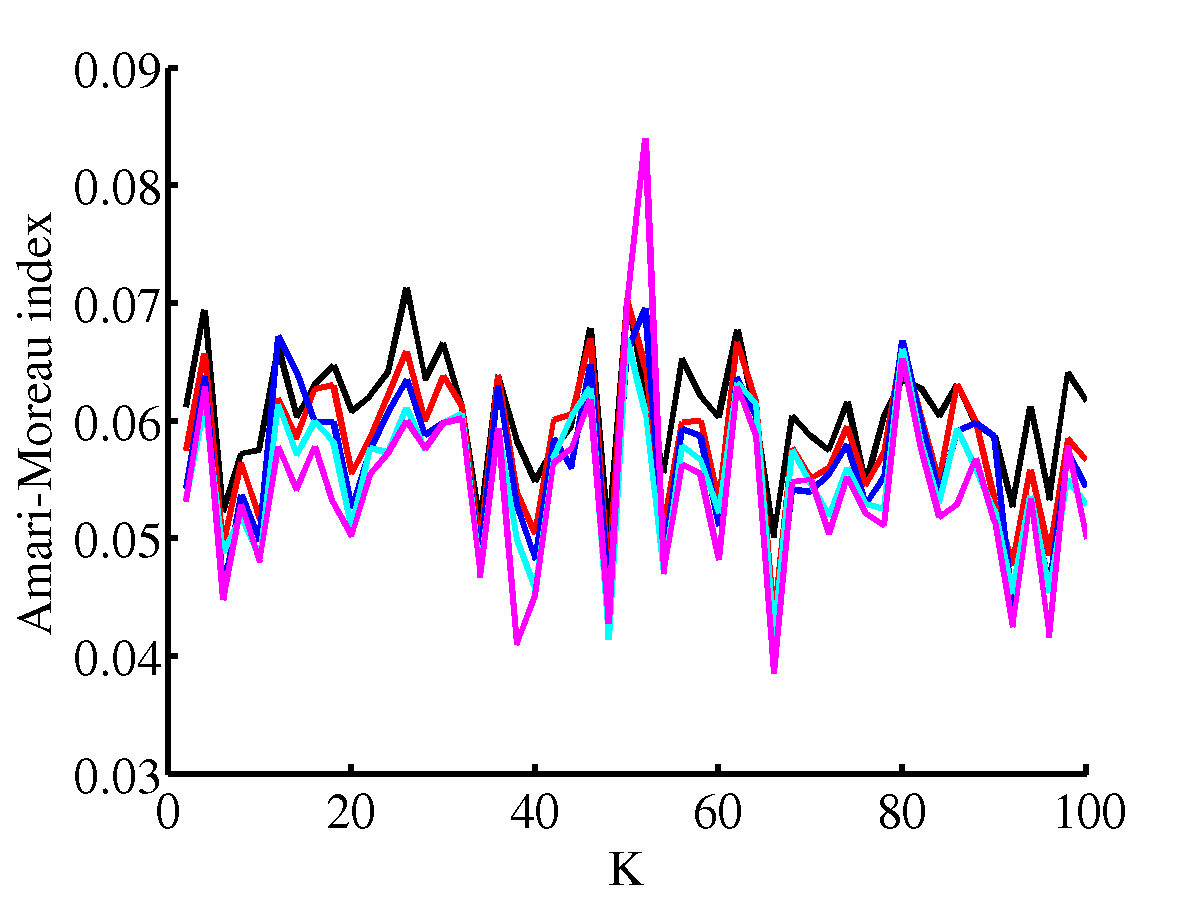}}
\end{figure}
\end{center}

In what follows, we report the Amari-Moreau performance indices using $K=100$ simulations, by generating different data matrices for every experiment, using the {\tt Jadiag}, {\tt Uwedge} and the {\tt LD-Newton} algorithms. For more complexity, we change the size $N \in \{ 50, 100 \}$ of the matrices to be approximately joint diagonalized and we increase the noise level up to 0.1. Finally, to get a statistical comparison of the Amari-Moreau indices obtained by the three algorithms, we give the means (averages) for every method through the 100 simulations. Results for different values of $\alpha \in \{ -0.75, 0, 0.75 \}$ are presented in Fig.~\ref{fig:ami-K} and Tab.~\ref{tab1}.

\begin{center}
\begin{table}[!htbp]
\centering
\caption{The mean (average) for $100$ Amari-Moreau indices obtained from 100 different simulations using the {\tt Jadiag}, {\tt Uwedge} and {\tt LD-Newton} (with different values of the parameter $\alpha$) algorithms for data with $SNR=0.1$.}
\label{tab1}
\begin{tabular}{|l|l|l|l|l|}
\hline
{\bf Method} & {\tt Jadiag}            & {\tt Uwedge}      & \multicolumn{2}{c|}{\tt LD-Newton} \\ \hline
\multirow{3}{*}{$N=20$} & \multirow{3}{*}{0.1162} & \multirow{3}{*}{0.1284} & $\alpha=-0.75$         & 0.1173        \\ \cline{4-5} 
                        &&                         & $\alpha=0$             & 0.1141        \\ \cline{4-5} 
                        &&                         & $\alpha=0.75$          & 0.1108        \\ \hline
\multirow{3}{*}{$N=50$} & \multirow{3}{*}{0.0611} & \multirow{3}{*}{0.0591} & $\alpha=-0.75$         & 0.0574        \\ \cline{4-5} 
                        &&                         & $\alpha=0$             & 0.0557        \\ \cline{4-5} 
                        &&                         & $\alpha=0.75$          & 0.0548        \\ \hline
\end{tabular}
\end{table}
\end{center}

The numerical results of the simulations show that these methods are convergent. Even though the {\tt Jadiag} and the {\tt Uwedge} algorithms have the faster convergence in terms of the iterations number and computation time compared with the {\tt LD-Newton}, the latter has proven to have better performance and more accurate results in estimating the approximate joint diagonalizer for the set of data matrices (\ref{eq:data}). Based on the computations of the Amari-Moreau index, Figs.~\ref{fig:ami-iter} and \ref{fig:ami-K} show that the {\tt LD-Newton} algorithm has a better robustness for higher noise levels and larger data matrices than the {\tt Jadiag} and the {\tt Uwedge} algorithms. Furthermore, by varying the value of the parameter $\alpha  \in \{-0.75, 0, 0.75 \}$, through different simulations, the efficiency of the {\tt LD-Newton} algorithm has demonstrated a slight improvement if we increase the value of $\alpha$. From the statistical analysis of the  three algorithms, illustrated in Tab.~\ref{tab1}, the superiority of the {\tt LD-Newton} algorithm, for different value of $\alpha$, over other algorithms is fairly remarkable.

\section{Conclusion} \label{sec:conc}

We have introduced several properly-invariant measures for quantifying the closeness of a Hermitian positive-definite matrix to diagonality. These measures are in fact distances or divergences between the given positive-definite matrix and its diagonal part. We gave explicit expressions of these diagonality measures and discussed their invariance properties. We have also defined ``true"  diagonality measures as distances or divergences between the given positive-definite matrix and its closest diagonal positive-definite matrix. We have used the log-determinant $\alpha$-measure for the approximate joint diagonalization of a set of Hermitian positive-definite matrices. We described a modified Newton method for the numerical solution of the approximate joint diagonalization problem. The preliminary numerical results show that this method is convergent and has better performance properties than two other algorithms. Convergence analysis and further numerical investigations with real data will be the subject of a future work. 

To the best of our knowledge, our algorithm for $\alpha = 0$ is the only geometry-aware approximate joint diagonalizer algorithm using a symmetric function. For instance, the criterion used by Pham's algorithm, which is based on the Kullback-Leibler divergence, is not symmetric. As suggested in \cite{congedo15}, the geometric mean computed by an approximate joint diagonalizer approach using our $\alpha$-divergence would respect all desirable properties.

\appendix

\section{Useful relations for deriving the gradient and Hessian of the cost function} 

For $\bd{Z} \in \mathcal{M}_{m, n}(\mathbb{C})$, the vector $\vec{(\bd{Z}^T)} \in \mathbb{C}^{m n}$ is linearly related to the vector $\vec{(\bd{Z})} \in \mathbb{C}^{m n}$. This relation uniquely defines the $m n \times m n$ permutation matrix $\bd{P}_{m \times n}$ (with entries 0 or 1) through
\begin{equation} \label{vec1}
\vec{(\bd{Z}^T)} = \bd{P}_{m \times n} \vec{(\bd{Z})}.
\end{equation}
The inverse, $\bd{P}_{m \times n}^{-1}$, of $\bd{P}_{m \times n}$ satisfies $\bd{P}_{m \times n}^{-1} = \bd{P}_{m \times n}^T = \bd{P}_{n \times m}$. For the special case where $m=n$, the commutation matrix is an involution, i.e., $\bd{P}_{m \times m}^2 = \bd{I}_{m^2}$. 

Similarly, for $\bd{Z} \in \mathcal{M}_{m}(\mathbb{C})$, the vector $\vec(\Diag \bd{Z}) \in \mathbb{C}^{m^2}$ is linearly related to the vector $\vec(\bd{Z}) \in \mathbb{C}^{m^2}$. This relation defines the diagonal projection matrix $\bd{P}_{\Diag}$ through
\begin{equation} \label{Pdiag}
\vec(\Diag \bd{Z}) = \bd{P}_{\Diag} \vec{(\bd{Z})}.
\end{equation}

Let $\bd{Z}$, $\bd{W} \in \mathcal{M}_{m, n}(\mathbb{C})$ and $\bd{A}$, $\bd{B} \in \mathcal{M}_{n, m}(\mathbb{C})$. Then, the following relations hold:
\begin{align}
\tr(\bd{Z} \bd{A} \bd{W} \bd{B}) &= \vec{(\bd{Z})}^T \bd{P}_{M\times N}^T(\bd{B}^T\otimes \bd{A}) \vec{(\bd{W})}, \label{trz1} \\
\tr(\bd{Z}^H \bd{A} \bd{W} \bd{B}) &= \vec{(\bd{Z})}^H (\bd{B}^T \otimes \bd{A})\vec{(\bd{W})}, \label{trz2} \\
\tr(\bd{Z}^H \bd{A} \bd{W}^H \bd{B}) &= \vec{(\bd{Z})}^H (\bd{B}^T \otimes \bd{A})\bd{P}_{m \times n} \vec{(\bd{W})}^*. \label{trz3}
\end{align}
For $\bd{A} \in \mathcal{M}_{p, q}(\mathbb{C})$, $\bd{B} \in \mathcal{M}_{r, s}(\mathbb{C})$, and $\bd{Z} \in \mathcal{M}_{q,r}(\mathbb{C})$ we have the relations
\begin{align}
\vec{(\bd{A} \bd{Z} \bd{B})} &= (\bd{B}^T\otimes \bd{A}) \vec{(\bd{Z})}, \label{vec2} \\
\bd{A} \otimes \bd{B} &= \bd{P}_{p \times r} (\bd{B} \otimes \bd{A}) \bd{P}_{s \times q}. \label{perm}
\end{align}
Finally, for square matrices $\bd{A}$ and $\bd{B}$ we have
\begin{equation} \label{eq:TrDiag}
\tr\left(\Diag(\bd{A})\Diag(\bd{B})\right) = \tr\left(\bd{A}\Diag(\bd{B})\right) = \tr\left(\Diag(\bd{A})\bd{B}\right).
\end{equation}

\section{Derivation of the gradient and Hessian of the cost function} \label{appenB}

Before we proceed in the derivation of the gradient and Hessian, we recall the following important result.
\begin{proposition}[\cite{Joho08}] Let $\bd{A}, \bd{B} \in \mathcal{M}_{n}(\mathbb{C})$. Then, for sufficiently small $t$, we have
\begin{equation} \label{prop}
\log \left(\det(\bd{I} + t \bd{A} + t^2 \bd{B})\right) = t \tr \bd{A} + t^2 \left(\tr \bd{B} - \tfrac{1}{2} \tr \bd{A}^2 \right) + O(t^3).
\end{equation}
\end{proposition}

\subsection{Gradient and Hessian of $\mathcal{J}^{(\alpha)}_{\bd{M}}(\cdot)$}

From (\ref{e1}) and after expansion we have
\begin{align*}
\mathcal{J}^{(\alpha)}_{\bd{M}}(\bd{C} + t \bd{Z}) &= \log \det\bigg(\frac{1-\alpha}{2} \bd{C} \bd{M}\bd{C}^H + \frac{1+\alpha}{2}\Diag(\bd{C} \bd{M}\bd{C}^H)\nonumber \\
& \qquad + \frac{1-\alpha}{2}(t(\bd{C} \bd{M} \bd{Z}^H + \bd{Z}\bd{M}\bd{C}^H)+t^2\bd{Z}\bd{M} \bd{Z}^H \big) \\
& \qquad + \frac{1+\alpha}{2}\Diag\big(t(\bd{C} \bd{M} \bd{Z}^H + \bd{Z}\bd{M}\bd{C}^H) + t^2 \bd{Z} \bd{M} \bd{Z}^H \big) \bigg),
\end{align*}
which we can write as
\begin{equation} \label{lgdt}
\mathcal{J}^{(\alpha)}_{\bd{M}}(\bd{C} + t \bd{Z}) = \mathcal{J}^{(\alpha)}_{\bd{M}}(\bd{C}) + \log \det \left(\bd{I} + t \bd{A}_{\alpha} + t^2 \bd{B}_{\alpha} \right),
\end{equation}
where, for the sake of simplicity of subsequent development, we have set
\begin{align*}
\bd{A}_{\alpha} & := (\bd{Q}_{\alpha})^{-1} \left(\tfrac{1-\alpha}{2}(\bd{C} \bd{M} \bd{Z}^H+\bd{Z}\bd{M}\bd{C}^H) + \tfrac{1+\alpha}{2}\Diag(\bd{C} \bd{M} \bd{Z}^H+\bd{Z}\bd{M}\bd{C}^H) \right), \\ 
\bd{B}_{\alpha} & := (\bd{Q}_{\alpha})^{-1} \left(\tfrac{1-\alpha}{2}\bd{Z}\bd{M} \bd{Z}^H+\tfrac{1+\alpha}{2}\Diag(\bd{Z}\bd{M} \bd{Z}^H) \right), 
\end{align*}
with $\bd{Q}_{\alpha}$ is defined by
\begin{equation} \label{eq:Qalpha}
\bd{Q}_{\alpha} := \frac{1-\alpha}{2} \bd{C} \bd{M} \bd{C}^H + \frac{1+\alpha}{2} \Diag(\bd{C} \bd{M} \bd{C}^H).
\end{equation}
We note that $\bd{Q}_{-1} = \bd{C} \bd{M} \bd{C}^H$ and $\bd{Q}_{1} = \Diag(\bd{C} \bd{M} \bd{C}^H)$. 

By applying (\ref{prop}) to (\ref{lgdt}) we obtain
\begin{equation}\label{dltterms}
\mathcal{J}^{(\alpha)}_{\bd{M}}(\bd{C} + t \bd{Z}) = \mathcal{J}^{(\alpha)}_{\bd{M}}(\bd{C}) + t \tr \bd{A}_{\alpha} +t^2 \left( \tr \bd{B}_{\alpha} - \tfrac{1}{2} \tr (\bd{A}_{\alpha})^2 \right) + O(t^3).
\end{equation}
Now, we need to compute the traces of $\bd{A}_{\alpha}$, $\bd{B}_{\alpha}$ and $(\bd{A}_{\alpha})^2$. Using (\ref{eq:TrDiag}) and after some calculations we have
\begin{equation}
\tr \bd{A}_{\alpha} = \mathfrak{Re} \left\{ \tr \left(\bd{Z}^H\left((1-\alpha)(\bd{Q}_{\alpha})^{-1} + (1+\alpha) \Diag((\bd{Q}_{\alpha})^{-1}) \right) \bd{C} \bd{M} \right) \right \}. \label{traa}
\end{equation}
From the expression of $\bd{B}_{\alpha}$ and by using (\ref{trz2}) and (\ref{eq:TrDiag}) we obtain
\begin{equation}
\tr(\bd{B}_{\alpha}) = \vec{(\bd{Z})}^H \bd{M}^T \otimes \left( \tfrac{1-\alpha}{2}(\bd{Q}_{\alpha})^{-1} + \tfrac{1+\alpha}{2} \Diag((\bd{Q}_{\alpha})^{-1}) \right) \vec{(\bd{Z})}. \label{trbb}
\end{equation}
The trace of $(\bd{A}_{\alpha})^2$ can be conveniently written as
\begin{equation} \label{tra2}
\tr (\bd{A}_{\alpha})^2 = \left(\tfrac{1-\alpha}{2}\right)^2 a + \left(\tfrac{1-\alpha^2}{2}\right) b + \left(\tfrac{1+\alpha}{2}\right)^2 c,
\end{equation}
where
\begin{align*}
a &= \tr\left(\bd{Z}^H(\bd{Q}_{\alpha})^{-1} \left[ \bd{C} \bd{M} \bd{Z}^H + 2 \bd{Z}\bd{M}\bd{C}^H \right] (\bd{Q}_{\alpha})^{-1}\bd{C} \bd{M}\right) \\
& \quad + \tr\left(\bd{Z}\bd{M}\bd{C}^H(\bd{Q}_{\alpha})^{-1}\bd{Z}\bd{M}\bd{C}^H(\bd{Q}_{\alpha})^{-1}\right), \\
b &= \tr\left(\bd{Z}^H(\bd{Q}_{\alpha})^{-1} \Diag \left(\bd{C} \bd{M} \bd{Z}^H + \bd{Z}\bd{M}\bd{C}^H \right) (\bd{Q}_{\alpha})^{-1}\bd{C} \bd{M} \right)\\
& \quad + \tr\left(\bd{Z}\bd{M}\bd{C}^H(\bd{Q}_{\alpha})^{-1} \Diag \left(\bd{C} \bd{M} \bd{Z}^H + \bd{Z}\bd{M}\bd{C}^H \right) (\bd{Q}_{\alpha})^{-1} \right), \\
c &= \tr\left((\bd{Q}_{\alpha})^{-1} \Diag(\bd{C} \bd{M} \bd{Z}^H)(\bd{Q}_{\alpha})^{-1} \left[\Diag(\bd{C} \bd{M} \bd{Z}^H + 2 \bd{Z}\bd{M}\bd{C}^H) \right] \right)\\
& \quad + \tr\left((\bd{Q}_{\alpha})^{-1}\Diag(\bd{Z}\bd{M}\bd{C}^H)(\bd{Q}_{\alpha})^{-1}\Diag(\bd{Z}\bd{M}\bd{C}^H)\right).
\end{align*}
Using (\ref{trz1}), (\ref{trz2}), (\ref{trz3}) and (\ref{vec1}) we get
\begin{align*}
a &= 2 \mathfrak{Re}\big \{\vec{(\bd{Z})}^T\bd{P}_{n \times n}^T\big[(\bd{M}\bd{C}^H(\bd{Q}_{\alpha})^{-1})^T\otimes(\bd{M}\bd{C}^H(\bd{Q}_{\alpha})^{-1}) \big]\vec{(\bd{Z})}\big \}\\
& \quad + 2\vec{(\bd{Z})}^H\big[(\bd{M}\bd{C}^H(\bd{Q}_{\alpha})^{-1}\bd{C} \bd{M})^T\otimes (\bd{Q}_{\alpha})^{-1} \big]\vec{(\bd{Z})}.
\end{align*}
Similarly, we obtain
\begin{align*}
b &= 2\vec{(\bd{Z})}^H\big[\bd{M}^T\bd{C}^T(\bd{Q}_{\alpha})^{-T}\otimes (\bd{Q}_{\alpha})^{-1}\big]\bd{P}_{\Diag}\big[\bd{C}^*\bd{M}^T\otimes \bd{I}_n\big]\vec{(\bd{Z})}\\
& \quad + 2\mathfrak{Re}\big \{ \vec{(\bd{Z})}^T\big[\bd{M}\bd{C}^H(\bd{Q}_{\alpha})^{-1}\otimes (\bd{Q}_{\alpha})^{-T}\big]\bd{P}_{\Diag}\bd{P}_{m\times n}\big[\bd{C}^*\bd{M}^T\otimes \bd{I}_n]\vec{(\bd{Z})}.
\end{align*}
Now by use of (\ref{eq:TrDiag}), we have 
\begin{align*}
c &= \tr\left(\bd{Z}^H\Diag\left((\bd{Q}_{\alpha})^{-1}\Diag(\bd{C} \bd{M} \bd{Z}^H + 2 \bd{Z}\bd{M}\bd{C}^H)(\bd{Q}_{\alpha})^{-1}\right)\bd{C} \bd{M}\right)\\
& \quad + \tr\left(\bd{Z}\bd{M}\bd{C}^H\Diag\left((\bd{Q}_{\alpha})^{-1}\Diag(\bd{Z}\bd{M}\bd{C}^H)(\bd{Q}_{\alpha})^{-1}\right)\right).
\end{align*}
With the help of (\ref{Pdiag}), (\ref{trz1}), (\ref{trz2}), (\ref{trz3}) and (\ref{vec2}) we obtain
\begin{align*}
c &= \vec{(\bd{Z})}^H\big[(\bd{C} \bd{M})^T\otimes \bd{I}_n\big]\vec{\left(\Diag\left((\bd{Q}_{\alpha})^{-1}\Diag(\bd{C} \bd{M} \bd{Z}^H)(\bd{Q}_{\alpha})^{-1}\right)\right)} \\
& \quad + 2\vec{(\bd{Z})}^H\big[(\bd{C} \bd{M})^T\otimes \bd{I}_n\big]\vec{\left(\Diag\left((\bd{Q}_{\alpha})^{-1}\Diag(\bd{Z}\bd{M}\bd{C}^H)(\bd{Q}_{\alpha})^{-1}\right)\right)}\\
& \quad + \vec{(\bd{Z})}^T\bd{P}_{n \times n}^T\big[\bd{I}_n\otimes \bd{M}\bd{C}^H\big]\vec{\left(\Diag\left((\bd{Q}_{\alpha})^{-1}\Diag(\bd{Z}\bd{M}\bd{C}^H)(\bd{Q}_{\alpha})^{-1}\right)\right)}\\
&= \vec{(\bd{Z})}^H\big[(\bd{C} \bd{M})^T\otimes \bd{I}_n\big]\bd{P}_{\Diag}\big[(\bd{Q}_{\alpha})^{-T}\otimes (\bd{Q}_{\alpha})^{-1}\big]\vec{\left(\Diag(\bd{C} \bd{M} \bd{Z}^H)\right)}\\
& \quad + 2\vec{(\bd{Z})}^H\big[(\bd{C} \bd{M})^T\otimes \bd{I}_n\big]\bd{P}_{\Diag}\big[(\bd{Q}_{\alpha})^{-T}\otimes (\bd{Q}_{\alpha})^{-1}\big]\vec{\left(\Diag(\bd{Z}\bd{M}\bd{C}^H)\right)}\\
& \quad + \vec{(\bd{Z})}^T\bd{P}_{n \times n}^T\big[\bd{I}_n\otimes \bd{M}\bd{C}^H\big]\bd{P}_{\Diag}\big[(\bd{Q}_{\alpha})^{-T}\otimes (\bd{Q}_{\alpha})^{-1}\big]\vec{\left(\Diag(\bd{Z}\bd{M}\bd{C}^H)\right)}.
\end{align*}
By making use of (\ref{vec1}) and (\ref{vec2}) we get
\begin{align*}
c &= \vec{(\bd{Z})}^H\big[(\bd{C} \bd{M})^T\otimes \bd{I}_n\big]\bd{P}_{\Diag}\big[(\bd{Q}_{\alpha})^{-T}\otimes (\bd{Q}_{\alpha})^{-1}\big]\bd{P}_{\Diag}\big[\bd{I}_n\otimes \bd{C} \bd{M}\big]\bd{P}_{n \times n}\vec{(\bd{Z})}^* \\
& \quad + 2\vec{(\bd{Z})}^H\big[(\bd{C} \bd{M})^T\otimes \bd{I}_n\big]\bd{P}_{\Diag}\big[(\bd{Q}_{\alpha})^{-T}\otimes (\bd{Q}_{\alpha})^{-1}\big]\bd{P}_{\Diag}\big[(\bd{M}\bd{C}^H)^T\otimes \bd{I}_n\big]\vec{(\bd{Z})}\nonumber\\
& \quad + \vec{(\bd{Z})}^T\bd{P}_{n \times n}^T\big[\bd{I}_n\otimes \bd{M}\bd{C}^H\big]\bd{P}_{\Diag}\big[(\bd{Q}_{\alpha})^{-T}\otimes (\bd{Q}_{\alpha})^{-1}\big]\bd{P}_{\Diag}\big[(\bd{M}\bd{C}^H)^T\otimes \bd{I}_n\big]\vec{(\bd{Z})}.\nonumber
\end{align*}
If we apply (\ref{perm}) in the above we find
\begin{align*}
c &= \vec{(\bd{Z})}^H\big[(\bd{C} \bd{M})^T\otimes \bd{I}_n\big]\bd{P}_{\Diag}\big[(\bd{Q}_{\alpha})^{-T}\otimes (\bd{Q}_{\alpha})^{-1}\big]\bd{P}_{\Diag}\bd{P}_{M\times M}\big[\bd{C} \bd{M}\otimes \bd{I}_n \big]\vec{(\bd{Z})}^*\\
& \quad + 2\vec{(\bd{Z})}^H\big[(\bd{C} \bd{M})^T\otimes \bd{I}_n\big]\bd{P}_{\Diag}\big[(\bd{Q}_{\alpha})^{-T}\otimes (\bd{Q}_{\alpha})^{-1}\big]\bd{P}_{\Diag}\big[(\bd{M}\bd{C}^H)^T\otimes \bd{I}_n\big]\vec{(\bd{Z})}\\
& \quad + \vec{(\bd{Z})}^T\big[\bd{M}\bd{C}^H\otimes \bd{I}_n\big]\bd{P}_{n\times n}\bd{P}_{\Diag}\big[(\bd{Q}_{\alpha})^{-T}\otimes (\bd{Q}_{\alpha})^{-1}\big]\bd{P}_{\Diag}\big[(\bd{M}\bd{C}^H)^T\otimes \bd{I}_n\big]\vec{(\bd{Z})}.
\end{align*}
Finally, since $\bd{P}_{\Diag}\bd{P}_{n\times n} = \bd{P}_{n\times n}\bd{P}_{\Diag} = \bd{P}$ we have
\begin{align*}
c &= 2\vec{(\bd{Z})}^H\big[(\bd{C} \bd{M})^T\otimes \bd{I}_n\big]\bd{P}_{\Diag}\big[(\bd{Q}_{\alpha})^{-T}\otimes (\bd{Q}_{\alpha})^{-1}\big]\bd{P}_{\Diag}\big[(\bd{M}\bd{C}^H)^T\otimes \bd{I}_n\big]\vec{(\bd{Z})}\\
& \quad + 2\mathfrak{Re}\big \{\vec{(\bd{Z})}^T\big[\bd{M}\bd{C}^H\otimes \bd{I}_n\big]\bd{P}\big[(\bd{Q}_{\alpha})^{-T}\otimes (\bd{Q}_{\alpha})^{-1}\big]\bd{P}_{\Diag}\big[(\bd{M}\bd{C}^H)^T\otimes \bd{I}_n\big]\vec{(\bd{Z})}\big \}.
\end{align*}
Now, by replacing the expressions of $a$, $b$ and $c$ in (\ref{tra2}) and after rearrangement we obtain
\begin{align}
\tr (\bd{A}_{\alpha})^2 &= 2\vec{(\bd{Z})}^H\bigg[\left(\frac{1-\alpha}{2}\right)^2\left[(\bd{M}\bd{C}^H(\bd{Q}_{\alpha})^{-1}\bd{C} \bd{M})^T\otimes (\bd{Q}_{\alpha})^{-1}\right]\nonumber\\
& \hspace{1cm}+\left(\frac{1-\alpha^2}{2}\right)\big[\bd{M}^T\bd{C}^T(\bd{Q}_{\alpha})^{-T}\otimes (\bd{Q}_{\alpha})^{-1}\big]\bd{P}_{\Diag}\big[\bd{C}^*\bd{M}^T\otimes \bd{I}_n\big]\nonumber\\
& \hspace{1cm}+\left(\frac{1+\alpha}{2}\right)^2\big[(\bd{C} \bd{M})^T\otimes \bd{I}_n\big]\bd{P}_{\Diag}\big[(\bd{Q}_{\alpha})^{-T}\otimes (\bd{Q}_{\alpha})^{-1}\big]\bd{U}\bigg]\vec{(\bd{Z})}\nonumber\\
& \hspace{0.4cm}+2\mathfrak{Re}\bigg\{\vec{(\bd{Z})}^T\bigg[\left(\frac{1-\alpha}{2}\right)^2\bd{P}_{n \times n}^T\big[(\bd{M}\bd{C}^H(\bd{Q}_{\alpha})^{-1})^T\otimes(\bd{M}\bd{C}^H(\bd{Q}_{\alpha})^{-1}) \big]\nonumber\\
& \hspace{1cm}+\left(\frac{1-\alpha^2}{2}\right)\big[\bd{M}\bd{C}^H(\bd{Q}_{\alpha})^{-1}\otimes (\bd{Q}_{\alpha})^{-T}\big]\bd{P}\big[\bd{C}^*\bd{M}^T\otimes \bd{I}_n]\nonumber\\
& \hspace{1cm}+\left(\frac{1+\alpha}{2}\right)^2\big[\bd{M}\bd{C}^H\otimes \bd{I}_n\big]\bd{P}\big[(\bd{Q}_{\alpha})^{-T}\otimes (\bd{Q}_{\alpha})^{-1}\big]\bd{U}\bigg]\vec{(\bd{Z})}\bigg \}, \label{tracare}
\end{align}
where 
\begin{equation}\label{whatisu}
\bd{U} = \bd{P}_{\Diag} \left[ (\bd{M}\bd{C}^H)^T \otimes \bd{I}_n \right].
\end{equation}

Inserting (\ref{traa}), (\ref{tracare}) and (\ref{trbb}) into (\ref{dltterms})  yields
\begin{align}
\mathcal{J}^{(\alpha)}_{\bd{M}}(\bd{C} + t \bd{Z})&=\mathcal{J}^{(\alpha)}_{\bd{M}}(\bd{C}) + t \mathfrak{Re}\left \{\tr\left(\bd{Z}^H\left((1-\alpha)(\bd{Q}_{\alpha})^{-1}\bd{C} \bd{M} + (1+\alpha)\Diag((\bd{Q}_{\alpha})^{-1})\bd{C} \bd{M}\right)\right)\right \} \nonumber \\
&\quad + t^2\vec{(\bd{Z})}^H\bigg(\tfrac{1-\alpha}{2}(\bd{M}^T\otimes (\bd{Q}_{\alpha})^{-1})+\tfrac{1+\alpha}{2}(\bd{M}^T\otimes \Diag( (\bd{Q}_{\alpha})^{-1})) \nonumber \\
& \quad -\left(\tfrac{1-\alpha}{2}\right)^2\left[(\bd{M}\bd{C}^H(\bd{Q}_{\alpha})^{-1}\bd{C} \bd{M})^T\otimes (\bd{Q}_{\alpha})^{-1}\right] \nonumber \\
& \quad - \left(\tfrac{1-\alpha^2}{2}\right)\big[\bd{M}^T\bd{C}^T(\bd{Q}_{\alpha})^{-T}\otimes (\bd{Q}_{\alpha})^{-1}\big]\bd{P}_{\Diag}\big[\bd{C}^*\bd{M}^T\otimes \bd{I}_n\big] \nonumber\\
& \quad - \left(\tfrac{1+\alpha}{2}\right)^2\big[(\bd{C} \bd{M})^T\otimes \bd{I}_n\big]\bd{P}_{\Diag}\big[(\bd{Q}_{\alpha})^{-T}\otimes (\bd{Q}_{\alpha})^{-1}\big]\bd{U}\bigg)\vec{(\bd{Z})} \nonumber \\
&\quad - t^2\mathfrak{Re}\bigg \{\vec{(\bd{Z})}^T\bigg[\left(\tfrac{1-\alpha}{2}\right)^2\bd{P}_{n \times n}^T\big[(\bd{M}\bd{C}^H(\bd{Q}_{\alpha})^{-1})^T\otimes(\bd{M}\bd{C}^H(\bd{Q}_{\alpha})^{-1}) \big]\nonumber\\
& \quad + \left(\tfrac{1-\alpha^2}{2}\right)\big[\bd{M}\bd{C}^H(\bd{Q}_{\alpha})^{-1}\otimes (\bd{Q}_{\alpha})^{-T}\big]\bd{P}\big[\bd{C}^*\bd{M}^T\otimes \bd{I}_n] \nonumber \\
& \quad + \left(\tfrac{1+\alpha}{2}\right)^2\big[\bd{M}\bd{C}^H\otimes \bd{I}_n\big]\bd{P}\big[(\bd{Q}_{\alpha})^{-T}\otimes (\bd{Q}_{\alpha})^{-1}\big]\bd{U}\bigg]\vec{(\bd{Z})}\bigg \} + O(t^3). \label{com}
\end{align}
By comparison between (\ref{com}) and the matrix form of the second-order Taylor expansion (\ref{taylor}), we conclude that the gradient $\bd{G}^{(\alpha)}_{\bd{M}}(\bd{C})$ and Hessians $\bd{H}^{(\alpha)}_{\bd{M}}(\bd{C})$ and $\bd{S}^{(\alpha)}_{\bd{M}}(\bd{C})$ of $\mathcal{J}^{(\alpha)}_{\bd{M}}(\cdot)$ evaluated at $\bd{C}$ are given by
\begin{align}
\bd{G}^{(\alpha)}_{\bd{M}}(\bd{C}) &= (1-\alpha)(\bd{Q}_{\alpha})^{-1}\bd{C} \bd{M}+(1+\alpha)\Diag((\bd{Q}_{\alpha})^{-1})\bd{C} \bd{M}, \label{eq:Gc} \\ 
\bd{H}^{(\alpha)}_{\bd{M}}(\bd{C}) &= (1-\alpha)(\bd{M}^T\otimes (\bd{Q}_{\alpha})^{-1})+(1+\alpha)(\bd{M}^T\otimes \Diag( (\bd{Q}_{\alpha})^{-1})) \nonumber \\
& \quad -\tfrac{(1-\alpha)^2}{2}\left[(\bd{M}\bd{C}^H(\bd{Q}_{\alpha})^{-1}\bd{C} \bd{M})^T\otimes (\bd{Q}_{\alpha})^{-1}\right] \nonumber\\
& \quad -(1-\alpha^2)\big[\bd{M}^T\bd{C}^T(\bd{Q}_{\alpha})^{-T}\otimes (\bd{Q}_{\alpha})^{-1}\big]\bd{P}_{\Diag}\big[\bd{C}^*\bd{M}^T\otimes \bd{I}_n\big] \nonumber\\
& \quad -\tfrac{(1+\alpha)^2}{2}\big[(\bd{C} \bd{M})^T\otimes \bd{I}_n\big]\bd{P}_{\Diag}\big[(\bd{Q}_{\alpha})^{-T}\otimes (\bd{Q}_{\alpha})^{-1}\big]\bd{U}, \label{eq:Hc}\\ 
\bd{S}^{(\alpha)}_{\bd{M}}(\bd{C}) &= -\tfrac{(1-\alpha)^2}{2}\bd{P}_{n \times n}^T\big[(\bd{M}\bd{C}^H(\bd{Q}_{\alpha})^{-1})^T\otimes(\bd{M}\bd{C}^H(\bd{Q}_{\alpha})^{-1}) \big] \nonumber \\
& \quad -(1-\alpha^2)\big[\bd{M}\bd{C}^H(\bd{Q}_{\alpha})^{-1}\otimes (\bd{Q}_{\alpha})^{-T}\big]\bd{P}_{\Diag}\bd{P}_{n\times n}\big[\bd{C}^*\bd{M}^T\otimes \bd{I}_n] \nonumber \\
& \quad -\tfrac{(1+\alpha)^2}{2}\big[\bd{M}\bd{C}^H\otimes\bd{I}_n\big]\bd{P}_{n\times n}\bd{P}_{\Diag}\big[(\bd{Q}_{\alpha})^{-T}\otimes (\bd{Q}_{\alpha})^{-1}\big]\bd{U}, \label{eq:Sc}
\end{align}
where $\bd{Q}_{\alpha}$ is defined in (\ref{eq:Qalpha}), $\bd{U}$ is defined in (\ref{whatisu}) and $\bd{P} = \bd{P}_{\Diag}\bd{P}_{n\times n} = \bd{P}_{n\times n}\bd{P}_{\Diag}$.

We note that Joho gave in \cite{Joho08} the expressions of both the gradient and Hessian of the two special cases of $\mathcal{J}^{(\alpha)}_{\bd{M}}(\bd{C})$:
\begin{align*}
\mathcal{J}^{(-1)}_{\bd{M}}(\bd{C}) &= \log \det(\bd{C} \bd{M}\bd{C}^H), \\
\mathcal{J}^{(1)}_{\bd{M}}(\bd{C}) &= \log \det \Diag(\bd{C} \bd{M}\bd{C}^H).
\end{align*}

\subsection{Gradient and Hessians of $\widetilde{\mathcal{J}}^{(\alpha)}_{\bd{M}}(\cdot)$}

From the definitions (\ref{digms}), (\ref{e1}) and the expressions (\ref{eq:Gc}), (\ref{eq:Hc}), (\ref{eq:Sc}), the gradient and Hessian of $\widetilde{\mathcal{J}}^{(\alpha)}_{\bd{M}}(\cdot)$ can be expressed as
\begin{align*}
\widetilde{\bd{G}}^{(\alpha)}_{\bd{M}}(\bd{C}) &= \tfrac{4}{1-\alpha^2} \left((1-\alpha) \left( (\bd{Q}_{\alpha})^{-1} - (\bd{Q}_{-1})^{-1} \right) + (1+\alpha)\Diag((\bd{Q}_{\alpha})^{-1} - (\bd{Q}_{-1})^{-1}) \right) \bd{C} \bd{M}, \\
\widetilde{\bd{H}}^{(\alpha)}_{\bd{M}}(\bd{C}) &= \tfrac{4}{1-\alpha^2}\bigg((1-\alpha)\left(\bd{M}^T\otimes (\bd{Q}_{\alpha})^{-1}\right)+(1+\alpha)\left(\bd{M}^T\otimes \Diag( (\bd{Q}_{\alpha})^{-1})\right)\\
&\hspace{1.1cm}-\tfrac{(1-\alpha)^2}{2}(\bd{M}\bd{C}^H(\bd{Q}_{\alpha})^{-1}\bd{C} \bd{M})^T\otimes (\bd{Q}_{\alpha})^{-1}\\
&\hspace{1.1cm}-(1-\alpha^2)\big[\bd{M}^T\bd{C}^T(\bd{Q}_{\alpha})^{-T}\otimes (\bd{Q}_{\alpha})^{-1}\big]\bd{P}_{\Diag}\big[\bd{C}^*\bd{M}^T\otimes \bd{I}_n\big]\\
&\hspace{1.1cm}-\tfrac{(1+\alpha)^2}{2}\big[(\bd{C} \bd{M})^T\otimes \bd{I}_n\big]\bd{P}_{\Diag}\big[(\bd{Q}_{\alpha})^{-T}\otimes (\bd{Q}_{\alpha})^{-1}\big]\bd{U}\bigg)\\
&-\tfrac{2}{1+\alpha}\left(2\left[\bd{M}^T\otimes (\bd{Q}_{-1})^{-1}\right]-2\left[\bd{M}^T\bd{C}^T \otimes \bd{I}_n\right]\left[(\bd{Q}_{-1})^{-T}\otimes (\bd{Q}_{-1})^{-1}\right]\left[\bd{C}^*\bd{M}^*\otimes \bd{I}_n\right]\right)\\
&-\tfrac{2}{1-\alpha}\left(2\left[\bd{M}^T\otimes (\bd{Q}_{1})^{-1}\right]-2\left[\bd{M}^T\bd{C}^T \otimes \bd{I}_n\right]\left[(\bd{Q}_{1})^{-T}\otimes (\bd{Q}_{1})^{-1}\right]\bd{P}_{\Diag}\left[\bd{C}^*\bd{M}^*\otimes \bd{I}_n\right]\right),
\end{align*}
and
\begin{align*}
\widetilde{\bd{S}}^{(\alpha)}_{\bd{M}}(\bd{C}) &= \tfrac{4}{1-\alpha^2}\bigg(-\tfrac{(1-\alpha)^2}{2}\bd{P}_{n \times n}^T\big[(\bd{M}\bd{C}^H(\bd{Q}_{\alpha})^{-1})^T\otimes(\bd{M}\bd{C}^H(\bd{Q}_{\alpha})^{-1}) \big]\\
&\hspace{2cm}-(1-\alpha^2)\big[\bd{M}\bd{C}^H(\bd{Q}_{\alpha})^{-1} \otimes (\bd{Q}_{\alpha})^{-T}\big]\bd{P}\big[\bd{C}^*\bd{M}^T\otimes \bd{I}_n]\\
&\hspace{2cm}-\tfrac{(1+\alpha)^2}{2}\big[\bd{M}\bd{C}^H \otimes \bd{I}_n\big]\bd{P}\big[(\bd{Q}_{\alpha})^{-T} \otimes (\bd{Q}_{\alpha})^{-1} \big]\bd{U}\bigg)\\
&\hspace{0.4cm}-\tfrac{2}{1+\alpha}\left(-2\left[\bd{M}\bd{C}^H\otimes \bd{I}_n\right]\left[(\bd{Q}_{-1})^{-1} \otimes (\bd{Q}_{-1})^{-T} \right]\bd{P}_{M\times M}\left[\bd{C}^*\bd{M}^*\otimes \bd{I}_n\right]\right)\\
&\hspace{0.4cm}-\tfrac{2}{1-\alpha}\left(-2\left[\bd{M}\bd{C}^H\otimes \bd{I}_n\right]\left[(\bd{Q}_{1})^{-1} \otimes (\bd{Q}_{1})^{-T} \right]\bd{P}\left[\bd{C}^*\bd{M}^*\otimes \bd{I}_n\right]\right).
\end{align*}

\bigskip
\noindent{\bf Acknowledgment.}
M.C. and M.M. would like to acknowledge partial support from the European project ERC-2012-AdG-320684-CHESS.

The authors are grateful for the excellent job of the anonymous referee, whose suggestions and remarks helped to improve the early version of this paper.

\section*{References}

\begin{thebibliography}{10}
\expandafter\ifx\csname url\endcsname\relax
  \def\url#1{\texttt{#1}}\fi
\expandafter\ifx\csname urlprefix\endcsname\relax\def\urlprefix{URL }\fi
\expandafter\ifx\csname href\endcsname\relax
  \def\href#1#2{#2} \def\path#1{#1}\fi

\bibitem{Flury86}
B.~Flury, W.~Gautschi, An algorithm for simultaneous orthogonal transformation
  of several positive definite symmetric matrices to nearly diagonal form, SIAM
  Journal on Scientific and Statistical Computing 7~(1) (1986) 169--184.

\bibitem{Flury84}
B.~N. Flury, Common principal components in $k$ groups, Journal of the American
  Statistical Association 79~(388) (1984) 892--898.

\bibitem{Comon10}
P.~Comon, J.~Jutten, Handbook of Blind Source Separation, Independent Component
  Analysis and Applications, Academic Press, Oxford, UK/Burlington, USA, 2010.

\bibitem{cardoso1993}
J.-F. Cardoso, A.~Soulomiac, Blind beamforming for non-Gaussian signals, Proc.
  Inst. Elect. Eng. F 140~(6) (1993) 362--370.

\bibitem{veen1998}
A.-J. van~der Veen, Algebraic methods for deterministic blind beamforming,
  Proceedings of the IEEE 86~(10) (1998) 1987--2008.

\bibitem{Joho08}
M.~Joho, Newton method for joint approximate diagonalization of positive
  definite {H}ermitian matrices, SIAM Journal on Matrix Analysis and
  Applications 30~(3) (2008) 1205--1218.

\bibitem{pham01}
D.~Pham, Joint approximate diagonalization of positive definite {H}ermitian
  matrices, SIAM Journal on Matrix Analysis and Applications 22~(4) (2001)
  1136--1152.

\bibitem{cardoso1996jacobi}
J.~F. Cardoso, A.~Souloumiac, Jacobi angles for simultaneous diagonalization,
  SIAM Journal on Matrix Analysis and Applications 17~(1) (1996) 161--164.

\bibitem{mesloub}
A.~Mesloub, K.~Abed-Meraim, A.~Belouchrani, A new algorithm for complex non-orthogonal joint diagonalization based on shear and {G}ivens rotations, IEEE
  Transactions on Signal Processing 62~(8) (2014) 1913--1925.

\bibitem{tichavsky09}
P.~Tichavsky, A.~Yeredor, Fast approximate joint diagonalization incorporating
  weight matrices, IEEE Transactions on Signal Processing 57~(3) (2009)
  878--891.

\bibitem{yeredor2002non}
A.~Yeredor, Non-orthogonal joint diagonalization in the least-squares sense
  with application in blind source separation, IEEE Transactions on Signal
  Processing 50~(7) (2002) 1545--1553.

\bibitem{yeredor2004approximate}
A.~Yeredor, A.~Ziehe, K.~R. M\"{u}ller, Approximate joint diagonalization using
  a natural gradient approach, in: Independent Component Analysis and Blind
  Signal Separation, Springer, 2004, pp. 89--96.

\bibitem{ziehe2004fast}
A.~Ziehe, P.~Laskov, K.-R. M\"{u}ller, A fast algorithm for joint
  diagonalization with non-orthogonal transformations and its application to
  blind source separation, The Journal of Machine Learning Research 5 (2004)
  777--800.

\bibitem{Bhatia00}
R.~Bhatia, Pinching, trimming, truncating and averaging of matrices, Amer.
  Math. Monthly 107~(7) (2000) 602--608.

\bibitem{Bhatia89}
R.~Bhatia, M.-D. Choi, C.~Davis, Comparing a matrix to its off-diagonal part,
  in: H.~Dym, S.~Goldberg, M.~A. Kaashoek, P.~Lancaster (Eds.), Operator
  Theory: Advances and Applications, Vol.~40, Birkh\"auser Basel, 1989, pp.
  151--164.

\bibitem{Moakher05}
M.~Moakher, A differential geometric approach to the geometric mean of
  symmetric positive-definite matrices, SIAM Journal on Matrix Analysis and
  Applications 26~(3) (2005) 735--747.

\bibitem{bhatia07}
R.~Bhatia, Positive Definite Matrices, Princeton University Press, New Jersey,
  2007.

\bibitem{terras}
A.~Terras, Harmonic Analysis on Symmetric Spaces and Applications II,
  Springer-Verlag, New York, 1988.

\bibitem{Amari12}
S.-i. Amari, Differential geometry derived from divergence functions:
  information geometric approach, in: M.~Deza, M.~Petitjean, K.~Markov (Eds.),
  Mathematics of Distances and Applications, 1st Edition, ITHEA, Sofia,
  Bulgaria, 2012, pp. 9--23.

\bibitem{Bregman67}
L.~M. Bregman, The relaxation method of finding the common point of convex sets
  and its application to the solution of problems in convex programming, {USSR}
  Computational Mathematics and Mathematical Physics 7~(3) (1967) 200--217.

\bibitem{Zhang04}
J.~Zhang, Divergence function, duality, and convex analysis, Neural Computation
  16~(1) (2004) 159--195.

\bibitem{Chebbi12}
Z.~Chebbi, M.~Moakher, Means of {H}ermitian positive-definite matrices based on
  the log-determinant $\alpha$-divergence function, Linear Algebra and its
  Applications 436~(7) (2012) 1872--1889.

\bibitem{amari09}
S.-i. Amari, Divergence, optimization and geometry, in: C.~S. Leung, M.~Lee,
  J.~H. Chan (Eds.), Neural Information Processing: Proceedings of the 16th
  International Conference (ICONIP 2009), Vol. 5863 of LNCS, Springer-Verlag,
  Berlin Heidelberg, 2009, pp. 185--193.

\bibitem{MoakherBatchelor06}
M.~Moakher, P.~G. Batchelor, Symmetric positive-definite matrices: From
  geometry to applications and visualization, in: J.~Weickert, H.~Hagen (Eds.),
  Visualization and Processing of Tensor Fields, Mathematics and Visualization,
  Springer Berlin Heidelberg, 2006, pp. 285--298.

\bibitem{Moakher09}
M.~Moakher, Divergence measures and means of symmetric positive-definite
  matrices, in: D.~H. Laidlaw, A.~Vilanova (Eds.), New Developments in the
  Visualization and Processing of Tensor Fields, Mathematics and Visualization,
  Springer Berlin Heidelberg, 2012, pp. 307--321.

\bibitem{Smith61}
O.~K. Smith, Eigenvalues of a symmetric $3 \times 3$ matrix, Commun. ACM 4~(4)
  (1961) 168.

\bibitem{Horst35}
P.~Horst, A method for determining the coefficients of a characteristic
  equation, Ann. Math. Statist. 6~(2) (1935) 83--84.

\bibitem{congedo15}
M.~Congedo, B.~Afsari, A.~Barachant, M.~Moakher, Approximate joint
  diagonalization and geometric mean of symmetric positive definite matrices,
  PLoS ONE 10 (2015) e0121423.

\bibitem{bhatia06}
R.~Bhatia, J.~Holbrook, Riemannian geometry and matrix geometric means, Linear
  Algebra Appl. 413 (2006) 594--618.

\bibitem{arsigny2007}
V.~Arsigny, P.~Fillard, X.~Pennec, N.~Ayache, Geometric means in a novel vector
  space structure on symmetric positive‐definite matrices, SIAM Journal on
  Matrix Analysis and Applications 29~(1) (2007) 328--347.

\bibitem{lim2012}
Y.~Lim, M.~P\'alfia, Matrix power means and the {K}archer mean, Journal of
  Functional Analysis 262~(4) (2012) 1498--1514.

\bibitem{palfia2016}
M.~P\'alfia, Operator means of probability measures and generalized karcher
  equations, Advances in Mathematics 289 (2016) 951--1007.

\bibitem{sra2015}
S.~Sra, Positive definite matrices and the {S}-divergence, Proc. Amer. Math.
  Soc. (2015) 12953.

\bibitem{manton2002optimization}
J.~H. Manton, Optimization algorithms exploiting unitary constraints, IEEE
  Transactions on Signal Processing 50~(3) (2002) 635--650.

\bibitem{congedo16}
M.~Congedo, R.~Phlypo, A.~Barachant, A fixed-point algorithm for estimating
  power means of positive definite matrices, in: Proceedings of the EUSIPCO
  Conference, Budapest, Hungary, 2016.

\bibitem{macchi93}
O.~Macchi, E.~Moreau, Self-adaptive source separation by direct and recursive
  networks, in: Proceedings of the International Conference on Digital Signal
  Processing, 1993, pp. 1154--1159.

\end{thebibliography}

\end{document}